\documentclass[journal]{IEEEtran}

\usepackage{graphicx}
\usepackage{citesort}
\usepackage{amssymb}
\usepackage{amsfonts}
\usepackage{amsmath}
\usepackage{epsfig}
\usepackage{color}
\usepackage{fancybox}
\usepackage{textcomp}
\usepackage{multirow}
\usepackage{setspa ce}
\usepackage{psfrag}
\usepackage{subfigure}

\usepackage[ruled,vlined,linesnumbered]{algorithm2e}
\usepackage{mathrsfs}

    \def\Complex{{\rm\rule[.23ex]{.03em}{1.1ex}\kern-.3em{C}}}

    \newcommand{\be}{\begin{equation}} \newcommand{\ee}{\end{equation}}
    \newcommand{\bea}{\begin{eqnarray}} \newcommand{\eea}{\end{eqnarray}}
    \newcommand{\benum}{\begin{enumerate}} \newcommand{\eenum}{\end{enumerate}}

        \newcommand{\qb}{{\bf b}}

        \newcommand{\qs}{{\bf s}}
        
        \newcommand{\qu}{{\bf u}}

        \newcommand{\qx}{{\bf x}}
        \newcommand{\qy}{{\bf y}}
        \newcommand{\qz}{{\bf z}}

        \newcommand{\qA}{{\bf A}}
        
        \newcommand{\qC}{{\bf C}}
        \newcommand{\qD}{{\bf D}}
        \newcommand{\qE}{{\bf E}}
        \newcommand{\qF}{{\bf F}}
        
        \newcommand{\qH}{{\bf H}}
        \newcommand{\qI}{{\bf I}}

        \newcommand{\qW}{{\bf W}}

        \newcommand{\qzero}{{\bf 0}}

        \newcommand{\tqH}{{\widetilde{\qH}}}

        \newcommand{\bbC}{{\mathbb C}}

        \newcommand{\calO}{{\mathcal O}}
        \newcommand{\calP}{{\mathcal P}}

        \newcommand{\calX}{{\mathcal X}}

        \newcommand{\diag}{{\sf diag}}
        
        \newcommand{\tr}{{\sf tr}}
        \newcommand{\Ex}{{\sf E}}

        \newcommand*{\argmin}{\operatornamewithlimits{argmin}\limits}

\begin{document}
\title{Finite-Alphabet Precoding for Massive MU-MIMO with Low-resolution DACs}

\author{~Chang-Jen~Wang,~Chao-Kai~Wen,~Shi~Jin,~and~Shang-Ho~(Lawrence)~Tsai

\thanks{{C.-J.~Wang} is with the Institute of Electrical Control Engineering, National Chiao Tung University, Hsinchu 30010, Taiwan,
Email: {\rm  dkman0988@gmail.com}.}
\thanks{{C.-K.~Wen} is with the Institute of Communications Engineering, National Sun Yat-sen University, Kaohsiung 80424, Taiwan,
Email: {\rm chaokai.wen@mail.nsysu.edu.tw}.}
\thanks{{S.~Jin} is with the National Mobile Communications Research Laboratory, Southeast University, Nanjing 210096, P. R. China, Email: {\rm  jinshi@seu.edu.cn}.}
\thanks{{S.-H.~Tsai} is with the department of Electrical Engineering, National Chiao Tung University, Hsinchu 30010, Taiwan,
Email: {\rm shanghot@mail.nctu.edu.tw}.}

\thanks{The source codes for the proposed precoding algorithms are available on
GitHub: {https://github.com/Wangchangjen/Matlab\_IDE}.}
}

\markboth{IEEE Transactions on Wireless Communications,~Vol.~XX, No.~XX, XXX~2018}
{Shell \MakeLowercase{\textit{et al.}}: Bare Demo of IEEEtran.cls for Journals}

\maketitle
\begin{abstract}
Massive multiuser multiple-input multiple-output (MU-MIMO) systems are expected to be the core technology in fifth-generation wireless systems because they significantly improve spectral efficiency. However, the requirement for a large number of radio frequency (RF) chains results in high hardware costs and power consumption, which obstruct the commercial deployment of massive MIMO systems. A potential solution is to use low-resolution digital-to-analog converters (DAC)/analog-to-digital converters for each antenna and RF chain. However, using low-resolution DACs at the transmit side directly limits the degree of freedom of output signals and thus poses a challenge to the precoding design. In this study, we develop efficient and universal algorithms for a downlink massive MU-MIMO system with finite-alphabet precodings. Our algorithms are developed based on the alternating direction method of multipliers (ADMM) framework. The original ADMM does not converge in a nonlinear discrete optimization problem. The primary cause of this problem is that the alternating (update) directions in
ADMM on one side are biased, and those on the other side are unbiased. By making the two updates consistent in an unbiased manner, we develop two algorithms called iterative discrete estimation (IDE) and IDE2: IDE demonstrates excellent performance and IDE2 possesses a significantly low computational complexity. Compared with state-of-the-art techniques, the proposed precoding algorithms present significant advantages in performance and computational complexity.
\end{abstract}


\section*{I. Introduction}

With the expansion of the Internet of Things and the increase in the data rate demand for mobile devices,
the requirement for wireless data rate continues to surge. Massive multiuser multiple-input multiple-output (MU-MIMO) systems are believed to be a key technology for reaching and surpassing the high data rate demand. This technology involves equipping a base station (BS) with a few hundreds of antennas in a centralized \cite{centralized1,centralized2} or distributed \cite{distributed}
manner to achieve a quasi-orthogonal channel vector between users and a BS. This system demonstrates several advantages, including improvement of network coverage and cell throughput and enhancement of user energy efficiency.

Although the benefits of using massive MU-MIMO systems in BS increase with the number of antennas, the requirement for a large number of radio frequency (RF) chains results in high hardware costs and power consumption, which obstruct the commercial deployment of massive MIMO system. A potential solution is the use of low-resolution digital-to-analog converters (DACs)/ analog-to-digital converters (ADCs) for each antenna and RF chain \cite{I4,DAC-power,DAC-power2,Energy-Efficiency}.
High-resolution ADC chains are the most power-hungry component on the receiver side. Thus, the hardware complexity and power consumption can be exponentially reduced by decreasing the resolution (in bits) of ADCs \cite{I4}. Although power expenditure is dominated by power amplifiers (PAs) \cite{Energy-Efficiency,DAC-power,DAC-power2} on the transmit side, the use of low-resolution DACs reduces variations in amplitude and allows the PAs to operate closer to saturation, thus increasing the efficiency of PAs.
The problems caused by low-resolution ADCs/DACs have stimulated many discussions
\cite{syn,Che1,Che2,Che3,Che4,Li-17TSP,Jacobsson-17TWireless,Liang-16JSAC,LowR1,Liang-16JSAC,Liang-16TCOM,Che3,Che4,det1,det2,det3,det4,Dong-17ComLetter,Zhang-16ComLetter,Kong-17TWireless,Kong-17ArXiv,Fan-15ComLetter,LowR3,LowR5,Liang-16TCOM,Liang-16JSAC,SQUID,R1,R2,SQUID,R4,R5,R6}.
Several contributions have been proposed for uplink systems with low-resolution ADCs, and related studies have considered multifold assessments, such as time/frequency synchronization \cite{syn}, channel estimation \cite{Che1,Che2,Che3,Che4,Li-17TSP,Jacobsson-17TWireless,Liang-16JSAC}, data detection \cite{LowR1,Jacobsson-17TWireless,Liang-16JSAC,Liang-16TCOM,Che3,Che4,det1,det2,det3,det4}, and related performance analyses \cite{Dong-17ComLetter,Zhang-16ComLetter,Kong-17TWireless,Kong-17ArXiv,LowR3,LowR5,Liang-16TCOM,Liang-16JSAC,Fan-15ComLetter}.
To date, only a small number of contributions \cite{SQUID,R1,R2,SQUID,R4,R5,R6} consider problems in downlink systems with low-resolution DACs, which piqued our interest.

In downlink massive MU-MIMO systems, BS transmits data to multiple independent user equipment (UE) simultaneously. Maximal
ratio transmission, zero-forcing (ZF), and Wiener filter (WF) precoders are commonly used to mitigate inter-user interference (IUI) caused by simultaneous transmission.
Using low-resolution DACs at the transmit side directly limits the degree of freedom of output signals.
A straightforward approach is to quantize the values of these conventional precoders directly.
However, this approach results in a significant performance loss when heavily quantized DACs are applied.

{\bf Relevant prior art}---The use of 1-bit DACs at the transmitter not only ensures constant-envelope (CE) signals in the input of power amplifiers but also minimizes the
energy consumption of a DAC itself. Therefore, massive MU-MIMO systems with 1-bit DACs have elicited much attention \cite{R1,R2,SQUID,R4,R5,R6}. {For example, a minimum mean-square error (MMSE) criterion precoder for a 1-bit MIMO downlink system with
higher-order modulation signals was proposed in \cite{R1}. This work breaks the myth that the 1-bit precoder is restricted to
QPSK signaling.} By using biconvex relaxation, \cite{R2} proposed 1-bit precoding
algorithms for massive MU-MIMO systems that demonstrate better error-rate performance than the ZF precoder directly followed by quantization. Moreover, \cite{R2} considered VLSI architectures that enable hundreds of antennas to serve tens of UE.
In contrast to \cite{R2} where the precoders are designed for 1-bit massive MIMO systems based on the MMSE criterion to
mitigate IUI, \cite{R4} changed the design criterion to the minimum bit error-rate (BER).
In \cite{R5}, the proposed precoder design considers the signal distortions
caused by 1-bit quantization at the transmitter and receiver. Aside from 1-bit DACs, \cite{SQUID} investigated the problem of downlink precoding with low-resolution DACs (e.g., 1--3 bits) at the BS by using Bussgang's theorem. \cite{R6} extended the work of \cite{SQUID} from cases with frequency-flat channels to those with frequency-selective channels.

Another type of hardware-aware precoding is the use of CE
precoding (e.g., \cite{A1,am,A3,cross_entropy,A5,A6}), which reduces the peak-to-average power ratio (PAPR) in the output signals\footnote{CE precoding has a perfectly constant envelope at the discrete-time domain but does not result in continuous-time transmit signals with a perfectly constant envelope.
However, compared with precoding methods, which result in large amplitude variations in the discrete-time domain, CE precoding results in continuous-time transmit signals that have significantly improved PAPR \cite{CEP}.} and thus decreases the linearity requirements at the BS. In CE precoding, the transmitted signals are
strictly limited by a fixed amplitude, and their phases are optimized to minimize IUI.
Phase rotation can be implemented by installing an analog or digital phase shifter (PS) in each antenna. The authors in \cite{A1,am,A3,cross_entropy} assumed that infinite-resolution PSs can generate any required phase. The design of infinite-resolution PSs leads to high hardware complexity and power consumption. Low-resolution PSs are therefore typically used in practice. When finite-resolution PSs are employed,
a straightforward approach that utilizes the quantized values of each continuous PS in a finite set also leads to a significant performance loss \cite{A5,A6}. Notably, the 1-bit DAC precoding problem can be considered as a special case of CE precoding, in which the phase
of the transmitted signal is limited to only four different values.

The mentioned architectures, such as low-resolution DACs, low-resolution PSs, or their hybrids \cite{Energy-Efficiency} can be energy-efficient. Given that these architectures result in finite alphabet signals, we call them finite-alphabet precodings. Thus far, existing finite-alphabet precodings are designed individually. Thus far, existing finite-alphabet precodings are designed individually. For example, most algorithms are designed specifically for the 1-bit DAC precoding problem, and extending these algorithms from 1-bit DAC precoding to general finite-alphabet precodings remains unjustified. In fact, all of the precoding problems mentioned above are related to a nonlinear least-squares (NLS) problem that attempts to minimize IUI (formed by a minimum Euclidean norm) with a finite-alphabet feasible set. The NLS problem is non-convex and difficult to solve explicitly. A common approach to address this optimization problem is to formulate it into a nonlinear integer (discrete) optimization problem and solve it using the branch-and-cut technique \cite{mitchell2002branch} (such as the sphere-decoding method \cite{SQUID}).
However, the worst-case complexity of the sphere-decoding method increases significantly with the problem dimensions. Therefore, this approach is un-suitable for massive MIMOs with a large number of antennas.
In \cite{TB_CEP}, the authors proposed an algorithm called trellis-based CE precoder (TB-CEP) that searches the precoding by using a trellis structure and
only retains a few possible combinations of the trellis states. However, TB-CEP does not provide a good trade-off between complexity and performance (which will be shown subsequently in our simulations).
Meanwhile, the alternating direction method of multipliers (ADMM) is a common algorithm for nonlinear discrete optimization. The individual steps in ADMM can be implemented exactly. However, ADMM does not converge, and even when it converges, it does not converge to a good suboptimal point.

{\bf Contribution}---Unlike existing algorithms that are individually designed for each specific architecture, we develop a \emph{universal} algorithm for a downlink massive MU-MIMO system with finite-alphabet precodings that minimize IUI. Most importantly, compared with existing algorithms, the proposed precoding algorithms also present significant advantages in performance and computational complexity. The key contributions of this study are threefold.

\begin{itemize}
\item Our algorithms are developed based on the ADMM framework. We reveal that the primary reason for the failure of ADMM to generate a good solution in nonlinear discrete optimization is as follows: the alternating (update) directions in ADMM on one side are biased and those on the other side are unbiased. By making the two updates consistent in a unbiased manner, we develop a universal framework for finite-alphabet precodings called iterative discrete estimation (IDE), which possesses better convergence properties than other local optimization methods. The proposed algorithms possess a unified structure, such that they can be applied to various finite-alphabet problems.

\item Although IDE generally achieves excellent error-rate performance, it has a slightly higher complexity than state-of-the-art methods because it requires matrix inversion. Following the same framework as IDE but using an approximation for matrix inversion, we propose a low-complexity version of IDE called IDE2. The simulations show that IDE2 provides good trade-offs between complexity and error-rate performance and is thus highly suitable for massive MU-MIMO systems.

\item Most authors (e.g., \cite{R2,SQUID,R6}) only evaluated their precoders under QPSK signaling. In fact, their precoders do not need to perform well under a generally high QAM signaling (e.g., 16-QAM and 64-QAM) because the precoders only need to transform the desired signals into four quadrants for QPSK signaling. In contrast to state-of-the-art precoders, IDE and IDE2 are universal and can work efficiently under any high-level QAM signaling.

\end{itemize}

The remainder of this paper is organized as follows. Section II introduces the system model and problem formulation. Section III presents ADMM and the reason for its failure to work well in nonlinear discrete optimization problems. The proposed algorithms are introduced, optimized, and simplified. Moreover, the complexity of the proposed algorithms is compared with the complexity of state-of-the-art methods. The simulations are presented in Sections IV. The conclusions are provided in Section V.

{\bf Notations}---For any matrix $\mathbf{A}$, $\mathbf{A}^{H}$ is the conjugate transpose of $\mathbf{A}$ and ${\sf tr}(\mathbf{A})$ denotes the traces of $\mathbf{A}$.
${\sf diag}(\qA)$ returns a diagonal matrix with its diagonal elements containing the diagonal elements of $\qA$. $\Ex_{x} \{ \cdot \}$ represents the expectation with respect to random variable $x$. When a complex-valued random variable $x$ is the Gaussian distribution with mean $\mu$ and variance $\sigma^2$, we write $x \sim \mathcal{CN}(\mu,\sigma^2)$.
\section*{II. System Model and Problem Formulation}

\subsection*{ A. System Model}
\begin{figure}\centering
    \resizebox{3.5in}{!}{%
    \includegraphics*{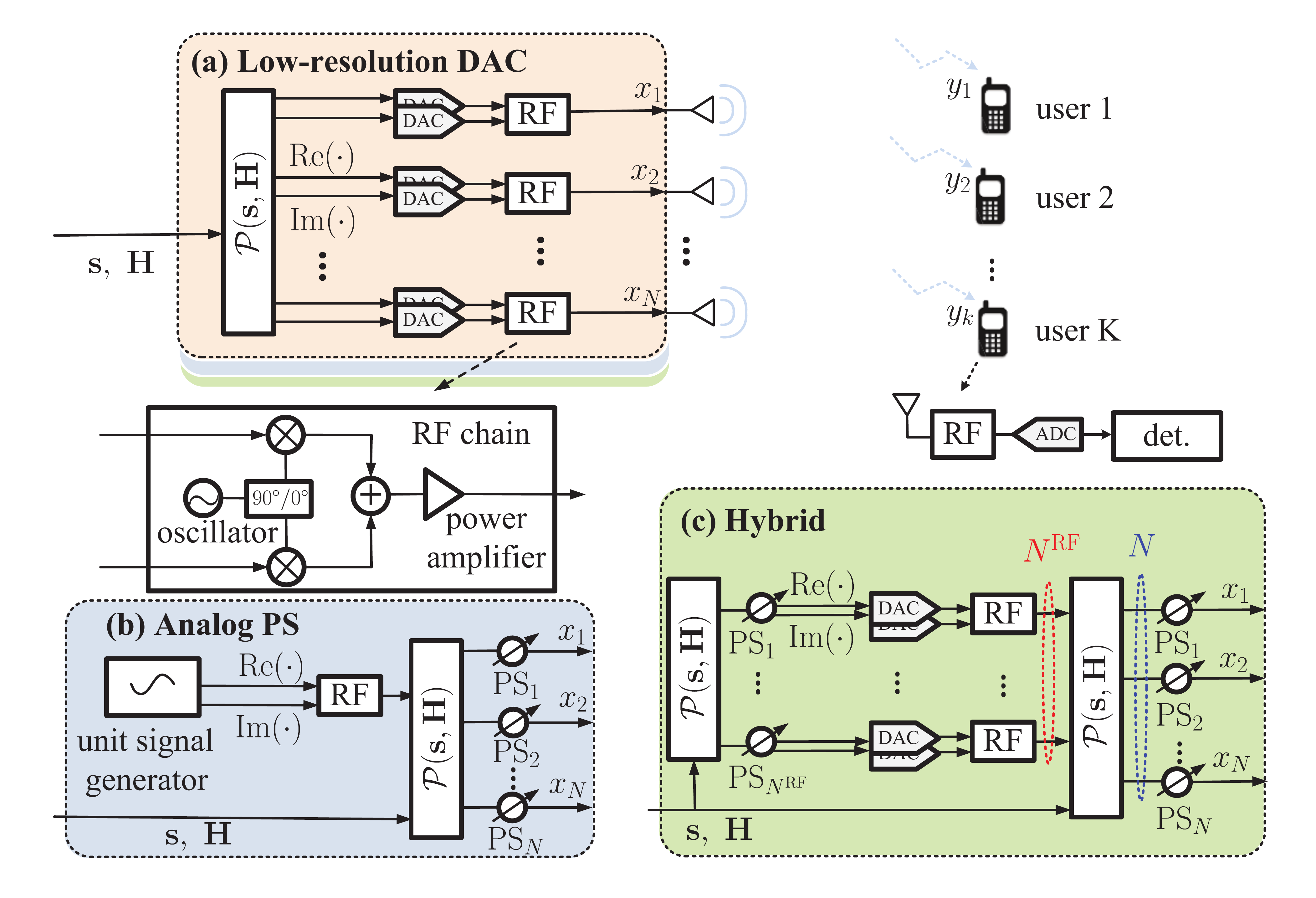} }%
    \caption{Massive MU-MIMO downlink system with the BS utilizing (a) low-resolution DACs, (b) low-resolution analog PSs, (c) hybrid. \label{Model}}
\end{figure}

As illustrated in Fig. \ref{Model}, we consider the downlink transmission of a massive MU-MIMO system, in which a BS with $N$ antennas serves $K$ single-antenna users simultaneously in the same time frequency resource. For simplicity, we assume that the ADCs at the UEs have infinite resolution.
The input output of the downlink frequency-flat fading channel can be expressed as\footnote{For the frequency-selective fading channel, the input-output of the downlink channel still can be expressed in a form similar to (1). Please see \cite[Eq. (10)]{R6} for details. As such, the proposed algorithms also can be applied to the frequency-selective fading channel.}
\begin{equation}\label{eq: in-out}
    \qy = \qH\qx+\qz,
\end{equation}
where $\qy=[y_1,\ldots,y_K]^{T}  \in\mathbb{C}^{K}$ contains the received signals of all users, $\qx=[x_1,\ldots,x_N]^{T} \in\mathbb{C}^{N}$ is the transmitted signal from the BS, $\qH = [H_{k,n}] \in\mathbb{C}^{K\times N}$ denotes the downlink channel with element $H_{k,n}$ being the channel response between transmitting antenna $n$ and user $k$, and $\qz=[z_1,\ldots,z_K]^{T} \in\mathbb{C}^{K}$ is the noise vector.
We assume that channel matrix $\qH$ is perfectly known at the BS and $z_k$'s are i.i.d. circularly symmetric complex Gaussian with mean $0$ and variance $\sigma^2$, that is, $\qz \sim \mathcal{CN}(0,\sigma^2 \qI)$.

In the downlink transmission, the BS aims to transmit constellation points $s_k \in \cal{O}$ for $k=1,\ldots,K$ to each of $K$ users, where $\cal{O}$ is the set of constellation points (e.g., QPSK, 16-QAM, and 64-QAM).
To this end, symbol vector $\qs = [s_1,\ldots,s_K]^{T}$ is mapped into $N$-dimensional vector $\qx$ through a precoder denoted by $\qx = \calP(\qs,\qH)$. The notation $\calP(\qs,\qH)$ implies that the precoder not only depends on the transmit constellation points $\qs$ but also utilizes the realization of channel matrix $\qH$. As an example, if the ZF precoder is used, then we have
\begin{equation}
    \qx = \calP (\qs,\qH) = \frac{1}{\beta} \qH^H(\qH\qH^H)^{-1} \qs,
\end{equation}
where $\beta$ is the precoding factor selected to ensure that a power constraint is satisfied. In this study, we consider the average power constraint as follows:
\begin{equation}\label{eq:power}
 \frac{1}{N}\Ex_{\qs}\{ \|\qx \|_2^2 \} \leq P_{\rm tx},
\end{equation}
where $P_{\rm tx}$ is the transmit power of each antenna. If $ \Ex_{\qs}\{ \qs\qs^H\} = \qI$, then we obtain $\beta = \sqrt{ \frac{\tr((\qH\qH^H)^{-1})}{N P_{\rm tx}} }$.
We define $ \mbox{SNR} = NP_{\rm tx}/\sigma^2$ as the signal-to-noise ratio (SNR). This definition of SNR includes the array gain and can be regarded as the received SNR perceived by each UE.

If the ZF precoder is used, then the received signal is $\qH\qx = \frac{1}{\beta} \qs$ rather than $\qs$. The users should (be able to) rescale the received signal by a factor $\beta$ to obtain an estimate of the transmit constellation points. Therefore, we define the metric for IUI as
\begin{equation} \label{eq:MUI}
{\rm IUI} = \Ex_{\qs}\{\| \qs - \beta \qH\qx\|_2^2 \}.
\end{equation}
Although the ZF precoder can achieve zero IUI, it presents several challenges to the BS, such as requiring infinite-resolution DACs and high-linearity power amplifiers.
In this study, we are interested in a practical setting where each antenna is equipped with a low-cost constrained RF chain.
For example, the BS is equipped with low-resolution DACs (e.g., 1-bit DACs), low-resolution analog PSs, or hybrid architectures, as illustrated in Figs. \ref{Model}(a), \ref{Model}(b), and \ref{Model}(c), respectively. The former two architectures are straightforward, and the hybrid architecture is highly general. To better understand the hybrid architecture, we consider an example that the number of DAC/RF chains is $N^{\rm RF}$, the resolution of DACs is one-bit, and each PS employs four discrete-phase resolution 4-PSK. In this case, the signal outputs of each complex ADC is one of the following $4$-QAM signals: $\{\pm 1 \pm j \}$. The output of each DAC/RF chain $i$ is then connected to $N$ PSs and yields the precoded signal $\qx_i$, which is 16-QAM that results from the superposition of $4$-QAM and 4-PSK. Afterwards, signals $\qx_i$ are combined by $N^{\rm RF}$-port power combiners to generate the final transmitted signal $\qx = \sum_{i=1}^{N^{\rm RF}} \qx_i$. This architecture employs in total $N^{\rm RF} N$ PS elements. Notice that $\qx$ can be represented by a finite set of values. Designing the finite-alphabet set can relax the linearity requirement, thus allowing the amplifiers to operate closer to saturation and increasing their efficiency.

In these applications, each entry of transmit vector $\qx$ is restricted to a finite-alphabet $\calX =\{\chi_0,\ldots,\chi_{M-1}\}$, where $\chi_m$ represents the possible quantization output.
We refer to ${M=|\calX|}$ and ${B=\log_2 M}$ as the number of quantization levels (per dimension\footnote{In low-resolution DAC case, we assume the same quantization alphabet for the real and imaginary parts. Therefore, the $n$th entry of the transmit vector $\qx$ is $x_n=x_{R,n} +j x_{I,n}$ with $x_{R,n}, x_{I,n} \in \calX$.}) and the number of quantization bits (per dimension), respectively.
We call the transmitted signal $\qx \in \calX^{N}$ the finite-alphabet precoder.

\subsection*{ B. Problem Formulation}
{If the precoder belongs to a finite-alphabet, that is, $\qx \in \calX^N$, then obtaining zero IUI becomes difficult in general.} The users will experience additional distortion.
Our goal is to design a precoder that minimizes IUI under the power constraint (\ref{eq:power}).
Notably, we have rescaled the received signal by the factor $\beta$. Therefore, given the transmitted symbol $\qs$, the mean squared error of the estimated
symbols at the receivers can be written as \cite{SQUID}
\begin{equation} \label{eq:objFun}
 \| \qs - \beta \qH\qx\|_2^2 + \beta^2 K \sigma^2.
\end{equation}
Factor $\beta$ serves as a trade-off between IUI and noise enhancement.
By using (\ref{eq:objFun}) as the objective function,
the precoder design can be formulated as follows:
\begin{equation} \label{eq:unqMMSE1}
\begin{aligned}
\min_{\qx, \, \beta} & \quad \left\| \qs - \beta\qH\qx \right\|^2_2 + \beta^2 K \sigma^2, \\
{\rm s.t.}& \quad \qx \in \calX^{N}, ~\beta > 0.
\end{aligned}
\end{equation}
If $\qx \in \calX^N$, then the average power of $\qx$ is always determined. Therefore, the power constraint is removed.

The problem formulation in (\ref{eq:unqMMSE1}) is general and can be used in several applications by setting support $\calX$ to be of several given forms.
For example, if the BS is equipped with infinite-resolution DACs, then we can set $\calX = \bbC$ and introduce the average power constraint \eqref{eq:power}. In this case,
the solution to \eqref{eq:unqMMSE1} is the WF precoder \cite{WFP} expressed as follows:
\begin{equation} \label{eq:wf}
    \qx = \frac{1}{\beta_{\rm WF}} \underbrace{\qH^H\left(\qH\qH^H + \frac{K \sigma^2}{N P_{\rm tx}} \qI \right)^{-1}}_{\triangleq \qW_{\rm WF}} \qs,
\end{equation}
where
\begin{equation*}
\beta_{\rm WF} = \sqrt{ \frac{\tr(\qW_{\rm WF}^H\qW_{\rm WF})}{ N P_{\rm tx}} }.
\end{equation*}
The 1-bit quantized precoder problem \cite{SQUID,R2} can be obtained by setting
$\calX=\sqrt{P_{\rm tx}}(\pm 1 \pm j) $.
If $\calX = \sqrt{P_{\rm tx}} \, e^{j \frac{2 \pi k}{M}}$ with $k=0,\ldots,M-1$, the precoder is called a CE precoder
\cite{am,cross_entropy,CEP2}. In the CE precoder, a high-efficiency power amplifier can be used because the antenna elements have the same output
amplitude. The problem of \eqref{eq:unqMMSE1} is also related to an integer programming problem, which is NP-hard in general.
Many techniques, such as sphere decoding \cite{SQUID}, NOMAD \cite{NOMAD}, and TB-CEP \cite{TB_CEP}, have been proposed to solve these problems.
However, the computation complexity of these techniques dramatically increases with the increase in BS antennas $N$.
Recent interest has shifted to the design of numerically efficient precoding methods suitable for massive MU-MIMO systems.

\section*{III. Algorithm}
The simultaneously optimizing $\qx$ and $\beta$ make the problem increasingly complicated. In the following discussion, we consider the case with a fixed $\beta$ and remove $\beta^2 K \sigma^2$ from the objective function. Specifically, we denote $\tqH = \beta\qH$ and consider the following optimization problem
\begin{equation} \label{eq:unqMMSE}
\begin{aligned}
&\min_{\qx } && \| \qs - \tqH\qx \|^2_2  , \\
&~{\rm s.t.} && \qx \in \calX^{N}.
\end{aligned}
\end{equation}
We develop a numerically efficient algorithm for solving problem \eqref{eq:unqMMSE}.
We begin by introducing a commonly used ADMM framework for nonconvex problems to explain why it fails to generate a good solution, and we describe in
detail our novel algorithms.

\subsection*{A. Why ADMM Fail}

\begin{figure}
    \centering
    \resizebox{3.5in}{!}{%
    \includegraphics*{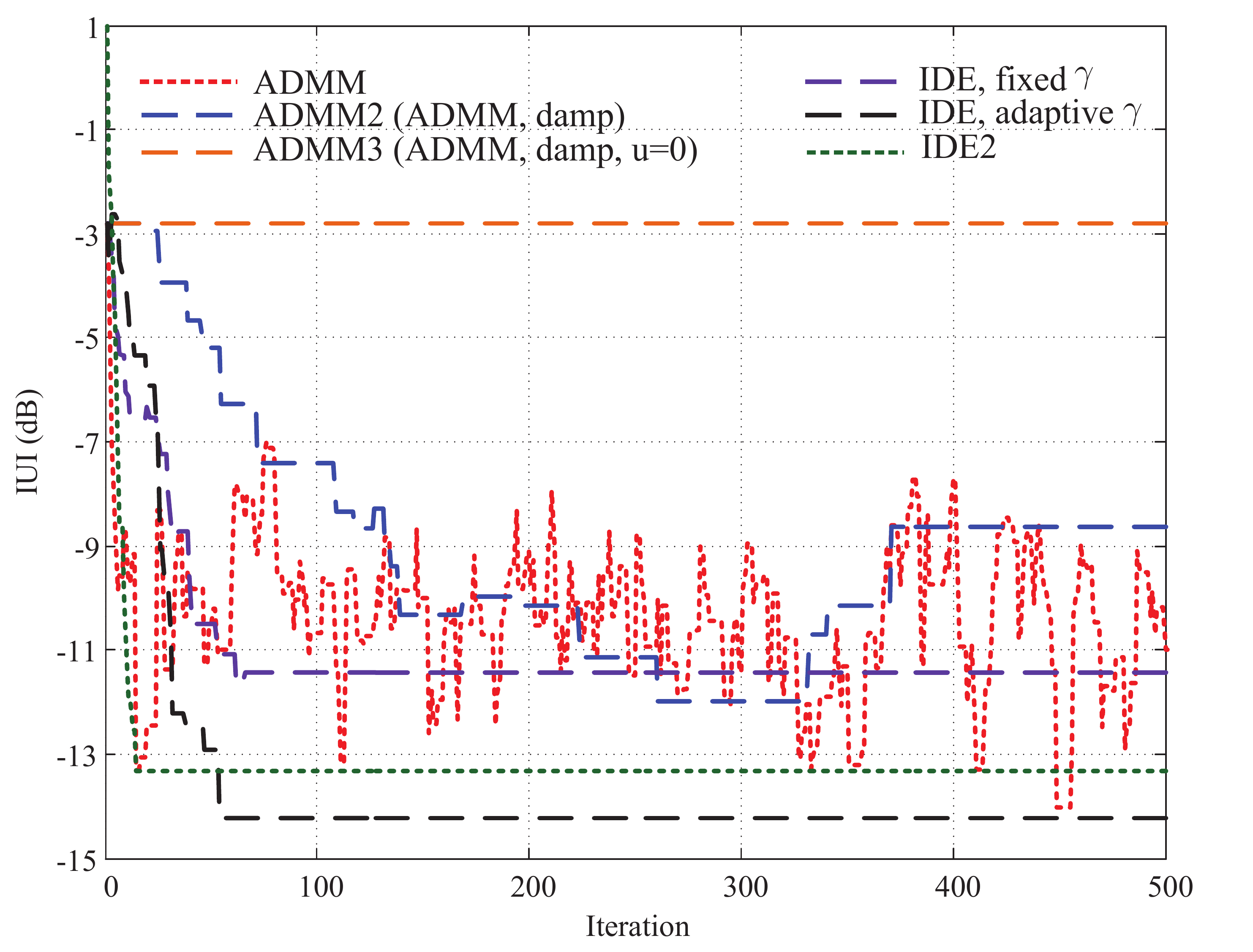} }%
    \caption{IUI versus the number of iterations in the same simulation environment for different algorithms.\label{F1}}
\end{figure}

To use the ADMM framework, we rewrite problem (\ref{eq:unqMMSE}) in a consensus form as follows \cite{ADMM}:
\begin{equation} \label{eq:unqMMSE2}
\begin{aligned}
&\min_{ \qx_1,\qx}&& \|\qs - \tqH\qx_1	\|_2^2 + I_{\calX}(\qx),\\
&~{\rm s.t.} && \qx_1 - \qx = 0,
\end{aligned}
\end{equation}
where $I_{\calX}(\cdot)$ is the indicator function of $\calX^N$, that is,
\begin{equation}
 I_{\calX}(\qx) = \left\{
 \begin{array}{ll}
 0, & \mbox{if } \qx \in \calX^N, \\
 \infty, & \mbox{otherwise}.
 \end{array}
 \right.
\end{equation}
The augmented Lagrangian of \eqref{eq:unqMMSE2} is expressed as
\begin{multline} \label{eq:Lr}
L_r\left(\qx_1,\qx,\qu\right) = \|\qs - \tqH\qx_1 \|^2_2 + I_{\calX}(\qx)
 \\ + \qu^{H} (\qx_1-\qx) +\gamma \left\|\qx_1-\qx \right\|^2_2,
\end{multline}
where $\qu$ is the dual vector, and $\gamma > 0$ is the penalty parameter (or the augmented Lagrangian parameter). The ADMM for this problem is expressed as
\begin{subequations} \label{eq:ADMM_1}
\begin{align}
\qx_1^{t+1} &=  \argmin_{\qx_1} L_r{\left( \qx_1,\qx^t, \qu^t \right)}, \\
\qx^{t+1} &=  \argmin_{\qx} L_r{\left( \qx_1^{t+1},\qx, \qu^t \right)} , \label{eq:qx22}\\
\qu^{t+1} &=  \qu^{t}+\gamma	{\left( \qx_1^{t+1}-\qx^{t+1} \right)}.
\end{align}
\end{subequations}
In \eqref{eq:ADMM_1}, the $\qx_1$-update involves solving the IUI minimization problem, the $\qx$-update involves projection onto a finite-alphabet set $\calX^{N}$, and the
$\qu$-update can be interpreted as a consensus adjustment step with step size $\gamma$. After some algebraic manipulation, \eqref{eq:ADMM_1} can be expressed explicitly as
\begin{subequations} \label{eq:ADMM_2}
\begin{align}
\qx_1^{t+1} &= \left(\tqH^H\tqH +\gamma\qI\right)^{-1}\left(\tqH^H\qs+\gamma\qx^t-\qu^t\right), \label{eq:ADMM_2_x1}\\
\qx^{t+1} &= \Pi_{\calX}{\left(\qx_1^{t+1} +\frac{1}{2\gamma}\qu^t \right)}, \label{eq:ADMM_2_x2} \\
\qu^{t+1} &= \qu^{t}+\gamma \left(	\qx_1^{t+1}-\qx^{t+1} \right), \label{eq:ADMM_2_u}
\end{align}
\end{subequations}
where $\Pi_{\calX}$ is projected onto $\{ x_n \in \calX, n =1,\ldots, N \}$.

The $\qx_1$-minimization step \eqref{eq:ADMM_2_x1} is convex, but the $\qx$-update \eqref{eq:ADMM_2_x2}
is projected onto a nonconvex set $\calX^{N}$. Although the use of ADMM for nonconvex problems is common, ADMM may not
converge. Fig. \ref{F1} shows the experimental result of IUI versus iteration for one trial, in which $N = 64$, $K = 16$, $H_{n,k} \sim \mathcal{CN}(0,1), \, \forall n,k$, and ${\calX = \frac{1}{\sqrt{128}} \{ \pm 1 \pm j \}}$, to better understand this problem. If we do not specify the update mechanism of $\gamma$, then the default is $\gamma = 1$ in subsequent experiments.
We see that the IUI of ADMM changes dramatically in each iteration and cannot converge.\footnote{We find that the same ADMM algorithm works well for massive MU-MIMO detection \cite{ADMM-infinity-norm}. Notice that the detection problem is completely different from the precoding problem at the $\qx_1$-update stage. For the detection problem, the $\qx_1$-update solves the least-squares solution, in which no local optimal point exists.
Therefore, the variation in each iteration is minor because the optimal solution $\qx$ is already close to that found through the least-squares solution, especially for a massive MIMO system.
However, for the precoding problem, the $\qx_1$-update itself has an infinitely number of solutions. Therefore, the variation in each iteration shall be too large for the iteration to converge.
} By checking the program itself, we determine that the main reason for this problem is projection $\Pi_{\calX}$.
In each iteration, projection $\Pi_{\calX}$ generates $\qx^{t+1}$ onto a discrete point, which makes $\qx^{t+1}$ clearly different from $\qx_1^{t+1}$, and the variation is too large to make the iteration converge.
To solve the rapid change problem, we introduce damping factor $\alpha \in [0,1]$ after \eqref{eq:ADMM_2_u} as follows:
\begin{subequations} \label{eq:ADMM_2-1}
\begin{align}
\qx_{\rm d}^{t+1} &= \alpha \qx_{\rm d}^{t} +\left(1-\alpha\right)\qx^{t+1},\\
\qu_{\rm d}^{t+1} &=\alpha \qu_{\rm d}^{t} + (1-\alpha)\qu^{t+1}.
\end{align}
\end{subequations}
We simply refer to ADMM \eqref{eq:ADMM_2} in conjunction with \eqref{eq:ADMM_2-1} as ADMM2. Notice that for ADMM2, $\qx^t$ in \eqref{eq:ADMM_2_x1} and $\qu^{t}$ in \eqref{eq:ADMM_2_u} should be replaced by $\qx_{\rm d}^{t}$ and $\qu_{\rm d}^{t}$, respectively. In Fig. \ref{F1}, ADMM2 is more stable than ADMM, and its track behaves as a stepped line because as the updates do not surpass a threshold, $\qx^{t+1}$ does not change in each iteration.

Although ADMM2 behaves like a smooth version of ADMM, the IUI track of ADMM2 is not \emph{monotonically} decreasing and still cannot converge.
By checking the program itself again, we realize that the problem is in the $\qu$-update. Recall that the $\qu$-update in ADMM is a consensus adjustment step, which attemps to make $\qx$ and $\qx_1$ reach a consensus.
However, this update cannot perform like a consensus adjustment as expected because the alternative update between $\qx_1$ and $\qx$ is conflictive: the $\qx$-update always outputs a discrete point in $\calX^{N}$, whereas the $\qx_1$-update never does. Trying to force $\qx_1$ and $\qx$ to reach a consensus instead results in divergence. Therefore, we remove dual vector $\qu$ in each iteration and obtain the following algorithm
\begin{subequations} \label{eq:ADMM_3}
\begin{align}
\qx_1^{t+1} &=  \left(\tqH^H\tqH +\gamma\qI\right)^{-1}\left(\tqH^H\qs+\gamma\qx_{\rm d}^t\right), \label{eq:ADMM_3_x1}\\
\qx^{t+1} &= \Pi_{\calX}{\left(\qx_1^{t+1} \right)}, \label{eq:ADMM_3_x2} \\
\qx_{\rm d}^{t+1} & = \alpha \qx_{\rm d}^{t} +\left(1-\alpha\right)\qx^{t+1}.
\end{align}
\end{subequations}
We refer to the algorithm as ADMM3. In ADMM3, $\qx_1$ and $\qx$ are updated in an alternating manner without a consensus adjustment.
Specifically, the $\qx_1$-update involves solving a linearly-constrained minimum Euclidean norm problem, i.e.,  
\begin{equation} \label{eq:linearConNorm}
   \rm  \argmin_{\qx_1} ~\|\qs - \tqH\qx_1 \|^2_2+\gamma \left\|\qx_1-\qx_{\rm d}^t \right\|^2_2.
\end{equation}
Then, the $\qx$-update projects the resulting point onto a finite-alphabet to obtain the subsequent iteration. This update strategy is straightforward. However, Fig. \ref{F1} shows that ADMM3
only updates one time and then falls into a local optimum solution with poor IUI.

\subsection*{B. Proposed Methods}

From the previous presented experiments, we realize that all ADMM-based algorithms experience a similar problem:
the $\qx_1$-update intends to minimize IUI, which is not necessarily an alphabet point, whereas the $\qx$-update aims to project the result onto an alphabet point.
Their updates cannot easily reach a consensus stage.
To understand this problem, we rewrite the $\qx_1$-update in \eqref{eq:ADMM_3_x1} as follows:
\begin{equation} \label{eq:bias_lmmse}
 \qx_1^{t+1} = \qx_{\rm d}^t + \qW {\left(\qs - \tqH\qx_{\rm d}^t\right)},
\end{equation}
where
\begin{equation} \label{eq:bias_w}
    \qW = {\left(\tqH^H\tqH +\gamma\qI\right)}^{-1}\tqH^H.
\end{equation}
From the perspective of estimation theory \cite{Unbiased}, \eqref{eq:bias_lmmse} can be interpreted as the optimal linear MMSE estimate of $\qx$ given prior knowledge on
\begin{equation} \label{eq:priorOfx}
\Ex\{ \qx \} = \qx_{\rm d}^t  ~~\mbox{and}~~ \Ex{\left\{ (\qx-\qx_{\rm d}^t) (\qx-\qx_{\rm d}^t)^H \right\}} = \frac{1}{\gamma} \qI.
\end{equation}
Such a linear MMSE estimate is \emph{biased} for each iteration  \cite{Unbiased} (see Appendix A for this argument).
However, the projection step in the $\qx$-update always returns an alphabet point, which is an \emph{unbiased} estimate of $\qx$.

To make the two updates consistent, we change \eqref{eq:bias_lmmse} into an unbiased version by
replacing $\qW$ with $\qW_{\rm u} = \qD \qW $, where $\qD$ is a diagonal matrix with its entries selected, such that the diagonal elements of $\qW_{\rm u} \tqH = \qD {(\tqH^H\tqH +\gamma\qI)}^{-1}\tqH^H \tqH$ are $1$.
If the diagonal elements of $\qW_{\rm u} \tqH$ are $1$, then the bias in \eqref{eq:bias_lmmse} is \emph{approximately} removed (see Appendix A for this argument).
To this end, we let $\qD = [\diag(\qW\tqH)]^{-1} $.
By substituting the unbiased version of \eqref{eq:bias_lmmse} into \eqref{eq:ADMM_3_x2}, we obtain the following algorithm
\begin{subequations} \label{eq:Proposed1}
\begin{align}
\qx^{t+1} &= \Pi_{\calX}{\left( \qx_{\rm d}^t + \qW_{\rm u} {\left(\qs - \tqH\qx_{\rm d}^t\right)} \right)},  \label{eq:Proposed1-1} \\
\qx_{\rm d}^{t+1} &=\alpha \qx_{\rm d}^{t} +\left(1-\alpha\right)\qx^{t+1}, \label{eq:Proposed1_1}
\end{align}
\end{subequations}
where
\begin{equation} \label{eq:Wu}
    \qW_{\rm u} = [\diag(\qW\tqH)]^{-1} \qW.
\end{equation}
Notably, we consider the diagonal entries before performing the matrix inversion in \eqref{eq:Wu}.
Given that each iteration consists of an estimation step and a discrete projection step, we refer to the algorithm as IDE.
Fig. \ref{F1} shows that the performance of IDE is significantly better than that of ADMM3 and does not have an instability problem similar to that of ADMM2.

The estimation in IDE is based on the problem in \eqref{eq:linearConNorm}, in which penalty parameter $\gamma$ is used to regulate IUI minimization and the previous estimate.
Generally, small values of $\gamma$ tend to produce a small IUI but at the expense of a low convergence rate.
Therefore, we use different penalty parameters $\gamma^{t}$ for each iteration with the goal of making performance less dependent on the choice
of the penalty parameter. Our setting of $\gamma^{t}$ is based on a simple observation: from linear estimation theory,
$\gamma$ in \eqref{eq:bias_w} should be set as an inverse of covariance \eqref{eq:priorOfx}, such that the estimate \eqref{eq:bias_lmmse} can achieve MMSE.
Given that the targeted $\qx$ is unknown, we estimate the error variance as follows:
\begin{equation} \label{eq:gammaSet}
 {(\gamma^{t})}^{-1} = \frac{\|\qs-\tqH\qx^{t} \|^2_2}{\tr(\tqH^H\tqH)}.
\end{equation}
In \eqref{eq:gammaSet}, $\tr(\tqH^H\tqH)$ serves as a normalization factor to remove the channel effect; thus, ${(\gamma^{t})}^{-1} \approx \frac{1}{N} \|\qx - \qx^{t} \|^2$ corresponds to the error variance \eqref{eq:priorOfx}.
The algorithm of IDE with adaptive $\gamma$ is summarized in Algorithm 1. We \emph{always} use IDE with adaptive $\gamma$; therefore, we simply refer to Algorithm 1 as IDE. Selection of damping factor $\alpha$ requires a trade-off between stability and speed of convergence. In Algorithm 1 (as well as Algorithm 2), we set $ \alpha= 0.95$ based on experience.
In addition, through exhaustive simulations, we find that the IUI results of the converged solutions with different initial points of $\qx_{\rm d}^{0}$ are similar. Therefore, we simply set $\qx_{\rm d}^{0}=\boldsymbol{0}$ as the initial point.

\begin{algorithm}[t]
 \textbf{Inputs}: $\qs$, $\tqH = \beta\qH$ \\
 \textbf{Initial}: $t=0$, $\qx_{\rm d}^0 = \textbf{0}$, $\gamma_{\rm d}^{0}= 1$,\ $\alpha = 0.95$ \\
 \While{$t < T$}{
 $\qW = {\left(\tqH^H\tqH +\gamma^{t} \qI\right)}^{-1}\tqH^H$ \\
 $\qD = [\diag(\qW\tqH)]^{-1} $  \\
 $\qx^{t+1} = \Pi_{\calX}{\left( \qx_{\rm d}^t + \qD \qW {\left(\qs - \tqH\qx_{\rm d}^t\right)} \right)} $\\
 $\gamma^{t+1} = \frac{\tr(\tqH^H\tqH)}{\|\qs-\tqH\qx^{t+1} \|^2_2}$\\
 $\qx_{\rm d}^{t+1} =\alpha \qx_{\rm d}^{t} +(1-\alpha)\qx^{t+1}$\\
 $\gamma_{\rm d}^{t+1}  =\alpha \gamma_{\rm d}^t +(1-\alpha)\gamma^{t+1}$\\
 $t \leftarrow t+1$}
 \textbf{Output}  $ \qx =\qx^{t+1}$
 \caption{IDE}
\end{algorithm}

Fig. \ref{F1} shows that IDE with adaptive $\gamma$ converges rapidly and performs significantly better than the other algorithms.
Although the performance of IDE is good, its complexity could be high for massive MIMO systems.
The complexity of IDE is dominated by the matrix inversion in line 4 of Algorithm 1, which is expressed as $\calO(NK^2)$.
Notably, matrix inversion has to calculate each iteration because $\gamma^{t}$ changes in each iteration.

We address the complexity issue by using an approximation for the matrix inversion. In particular, if $\gamma$ is large\footnote{From \eqref{eq:linearConNorm},
a large value of $\gamma$ implies $\qx_1 \approx \qx^t$; that is, $\qx_1$ is close to its previous iteration $\qx^t$. Therefore, we observe from Fig. \ref{F1} that IDE with this approximation exhibits fast convergence but high IUI.}, then we can obtain the following approximation
\begin{equation}
(\tqH^H\tqH+\gamma\qI)^{-1} \approx \frac{1}{\gamma} \qI,
\end{equation}
which results in $\qW \approx \frac{1}{\gamma} \tqH^{H}$. Consequently, we obtain
\begin{equation}
    \qW_{\rm u} \approx [\diag(\tqH^{H}\tqH)]^{-1} \tqH^{H}.
\end{equation}
By substituting $\qW_{\rm u}$ into  \eqref{eq:Proposed1}, we obtain a low-complexity version of IDE (summarized in Algorithm 2). We refer to the algorithm as IDE2.
As shown in Fig. \ref{F1}, the performance of IDE2 is only slightly degenerated because of the low complexity.

\begin{algorithm}[t]
 \textbf{Inputs}: $\qs$, $\tqH = \beta\qH$\\
 \textbf{Initial}: $t=0$, $\qx_{\rm d}^0 = \textbf{0}$, $\alpha = 0.95$\\
 \While{$t < T$}{
 $\qW_{\rm u} = [\diag(\tqH^{H}\tqH)]^{-1} \tqH^{H}$ \\
 $\qx^{t+1} = \Pi_{\calX}{\left( \qx_{\rm d}^t + \qW_{\rm u} {(\qs - \tqH\qx_{\rm d}^t)} \right)} $\\
 $\qx_{\rm d}^{t+1} = \alpha \qx_{\rm d}^{t} +(1-\alpha)\qx^{t+1}$\\
 $t \leftarrow t+1$}
 \textbf{Output} $ \qx =\qx^{t+1}$
 \caption{IDE2}
\end{algorithm}

IDE2 appears nearly similar to classical gradient descent. $F(\qx) = \| \qs - \tqH\qx \|_2^2$ denotes
the objective function of \eqref{eq:unqMMSE}. The negative gradient of $F(\qx)$ is $-\nabla F(\qx) = \tqH^{H} (\qs - \tqH\qx)$. In gradient descent, the update should be in the following form
\begin{equation}
\qx^{t+1} = \Pi_{\calX}{\left( \qx^t + \gamma \tqH^{H} {\left(\qs - \tqH\qx^t\right)} \right)},
\end{equation}
where step size $\gamma$ should be sufficiently small, such that $F(\qx^{t+1}) \leq F(\qx^{t})$. The main difference between IDE2 and gradient descent is the step size.
The set size in IDE2 is in vector form, i.e., $[\diag(\tqH^{H}\tqH)]^{-1}$, which attempts to make each update in an unbiased manner. Meanwhile, the set size in gradient descent
is a constant $\gamma$, which makes each update in a biased manner.
Following the similar principle above, we find that IDE with fixed $\gamma$ is similar to the classical Newton's method. However, in contrast to the relationship between gradient descent and IDE2 obtained through straightforward comparison, the relationship between Newton's method and IDE is more complex. We mainly show that adaptive $\gamma$ provides the best performance. Thus, a detailed discussion on this relationship is beyond the scope of this paper.

\subsection*{ C. Alternative Update Mechanism for $\beta$}

In the previous discussion, we fix precoding factor $\beta$. For any given $\qx^{t+1}$, precoding factor $\beta$ that minimizes \eqref{eq:unqMMSE1} is expressed as follows:
\begin{equation}\label{eq:alpha}
 \beta^{t+1} =\frac{\rm{Re}\left\{\qs^H\qH\qx^{t+1}\right\}}{ \|\qH\qx^{t+1}\|_2^2+K\sigma^2 }.
\end{equation}
Appendix B shows the derivation.
Then, we fix the estimate $\beta^{t+1}$ and use the proposed algorithms to derive $\qx^{t+2}$. The algorithm alternates between the updates of $\beta^{t}$ and $\qx^{t+1}$.
Specifically, we plug \eqref{eq:alpha} after line 6 of Algorithm 1 and line 5 of Algorithm 2 with initial $\beta^{0} = 1$.

\begin{table*}
\footnotesize
\centering
\caption{Computational complexity for different algorithms} \label{Table 1}
\begin{tabular}{|c|ccc|}
\hline
Algorithm & \multicolumn{1}{|c|}{1-st iteration}                  & \multicolumn{1}{|c|}{Subsequent iteration (each)}                &  \multicolumn{1}{c|}{$T$ iterations}   \\ \hline
\multicolumn{1}{|c|}{\multirow{3}{*}{SQUID}}     & \multicolumn{1}{|c|}{\multirow{2}{*}{ $2NK^2+\frac{1}{3}K^3$}} & \multicolumn{1}{|c|}{\multirow{3}{*}{ $2NK+N$}}&\multirow{2}{*}{ $T(2NK+N)+\frac{1}{3}K^3$} \\&\multicolumn{1}{|c|}{\multirow{3}{*}{ $+4NK+K^2+N$}}&&\multicolumn{1}{|c|}{\multirow{3}{*}{ $+2NK^2+2NK+K^2$}} \\&\multicolumn{1}{|c|}{}&&\multicolumn{1}{|c|}{}\\\hline
\multirow{3}{*}{C1PO}     & \multicolumn{1}{|c|}{\multirow{2}{*}{ $NK^2+\frac{1}{3}K^3$}} & \multicolumn{1}{|c|}{\multirow{3}{*}{ $2NK+K^2+N$}} &\multirow{2}{*}{ $T(2NK+K^2+N)+\frac{1}{3}K^3$} \\ &\multicolumn{1}{|c|}{\multirow{3}{*}{ $+2NK+K^2+2N$}}&&\multicolumn{1}{|c|}{\multirow{3}{*}{ $+NK^2$}}\\ &\multicolumn{1}{|c|}{}&\multicolumn{1}{|c|}{}&\\ \hline
\multirow{3}{*}{IDE}     & \multicolumn{1}{|c|}{\multirow{2}{*}{ $2NK^2+\frac{4}{3}K^3$}} & \multicolumn{1}{|c|}{\multirow{2}{*}{ $NK^2+\frac{4}{3}K^3$}} &\multirow{2}{*}{ $T(NK^2+\frac{4}{3}K^3+5NK$} \\ &\multicolumn{1}{|c|}{\multirow{3}{*}{ $+5NK+3N+K$}}&\multicolumn{1}{|c|}{\multirow{3}{*}{ $+5NK+3N+K$}}& \multirow{3}{*}{ $+3N+K)+NK^2$}\\ &\multicolumn{1}{|c|}{}&\multicolumn{1}{|c|}{}&\\ \hline
\multirow{2}{*}{IDE2}     & \multicolumn{1}{|c|}{\multirow{2}{*}{$4NK+3N$}} & \multicolumn{1}{|c|}{\multirow{2}{*}{$2NK+N$}} &\multirow{2}{*}{$T(2NK+N)+2NK+2N$} \\ &\multicolumn{1}{|c|}{}&\multicolumn{1}{|c|}{}&
\\ \hline\hline
\multirow{2}{*}{TB-CEP}    &                                                          & \multirow{2}{*}{$N^2KM^{L+1}$ }                        &                                          \\
   &                                                          &              &                                          \\
 \hline
\end{tabular}
\end{table*}

\begin{table*}
\footnotesize
\centering
\caption{The total number of multiplications (MCLs) for different algorithms}
\label{my-label}
\begin{tabular}{|c|c|c|c|}
\hline
                     &                          &                           &                         \\
\multirow{2}{*}{Algorithm}     & \multirow{2}{*}{Max iterations \#}                                  & \multirow{1}{*}{ MCLs \#}       & \multirow{1}{*}{ MCLs \#}     \\
                               &                                & \multirow{2}{*}{for $ (N,K)= (64,16)$}                               & \multirow{2}{*}{for $(N,K) = (128,16)$}   \\
                      &                          &                           &                         \\ \hline
 SQUID                               &  100                                             &  0.24E+6                 &   0.48E+6       \\ \hline
 C1PO                                &  24                                              & { \bf 0.07E+6}                 & {\bf  0.14E+6}                   \\ \hline
 IDE                                 &  100                                             &  2.7E+6                &  4.9E+6                   \\ \hline
IDE2                                & 100                                             & 0.21E+6                 & 0.43E+6               \\ \hline\hline
\multicolumn{1}{|l|}{} & \multicolumn{1}{l|}{} & \multicolumn{1}{l|}{} & \multicolumn{1}{l|}{} \\
\multirow{1}{*}{Output 4-PSK } & \multicolumn{1}{l|}{\multirow{1}{*}{$\ \ \ \ (N,K,L) $}} & \multicolumn{1}{l|}{$\ \ \ \ \ \ \ \ (N,K,L) $} & \multicolumn{1}{l|}{$\ \ \ \ \ \ (N,K,L) $}  \\
\multirow{2}{*}{($M=4$)} & \multicolumn{1}{l|}{\multirow{2}{*}{$\ =(64, 16, 0.1N)$}} & \multicolumn{1}{l|}{\multirow{2}{*}{$\ \ \ \ \ = (64, 16, 0.5N)$}} & \multicolumn{1}{l|}{\multirow{2}{*}{$\ \ \ = (64, 16, 0.9N)$}} \\ &&& \\ \hline
TB-CEP                             & 1000E+6                                          & 4.80E+24                                        & 2.00E+40                                        \\ \hline
\end{tabular}

\end{table*}

\subsection*{ D. Complexity Analysis}

We analyze the computational complexity of the proposed algorithms and other prior state-of-the-art methods, such as SQUID \cite{SQUID}, C1PO \cite{R2}, and TB-CEP \cite{TB_CEP}, in terms of the number of multiplication operations.
We only consider the real-valued model of \eqref{eq:unqMMSE}, that is, $\tqH$, $\qs$, and $\qx$ are real-valued matrices or vectors.
Our analytical results can be easily extended to the complex-valued model by using real-valued representation of the complex-valued matrix and vector.\footnote{A straightforward method is to replace the dimensions in the complexity analysis from $N$ and $K$ to $2M$ and $2K$, respectively.}

We analyze the complexity of IDE (Algorithm 1).
In line 4 of Algorithm 1, given that $\qH$ is fat (i.e., $K < N$), we apply the matrix inversion lemma \cite{inverse-lemma}
\begin{equation} \label{eq:matrixInvLemma}
(\tqH^H\tqH +\gamma^{t} \qI)^{-1} = \frac{1}{\gamma^t}{\left( \qI -\tqH^H \left(\tqH\tqH^H+ \gamma^t\qI \right)^{-1}\tqH \right)},
\end{equation}
and only compute the matrix inversion of the small matrix $\tqH\tqH^H + \gamma^{t} \qI$.
The computation of $(\tqH\tqH^H + \gamma^{t}\qI)^{-1}$ requires $ N K^2 + \frac{1}{3} K^3 $ multiplications, which involve the cost of forming $\tqH\tqH^H$ and computing the Cholesky factorization.
We cache $(\tqH\tqH^H +  \gamma^{t}\qI)^{-1}$ for the subsequent steps in lines 4 and 5.
Applying \eqref{eq:matrixInvLemma}, $\qW$ in line 4 is implemented by
\begin{multline}
 (\tqH^H\tqH +\gamma^{t} \qI)^{-1}\tqH^H \\
 = \frac{1}{\gamma^t} \left( \tqH^H -\tqH^H \left(  \tqH\tqH^H+ \gamma^t\qI \right)^{-1}\tqH\tqH^H \right).
\end{multline}
Notice that we do not compute $\qW$ in line 4 because it is the side product of the computation of $\qD$ in line 5.
Specifically, the computation of $\qD$ is implemented as follows:
\begin{multline}
 \left[\diag \left( \qW \tqH \right) \right]^{-1} = \Bigg[\frac{1}{\gamma^{t}} \cdot \\
 {\left( \diag(\tqH^H  \tqH) - \diag\Big(\tqH^H \Big(\tqH\tqH^H+ \gamma^t\qI\Big)^{-1} \tqH \tqH^H  \tqH\Big) \right)} \Bigg]^{-1},
\end{multline}
which involves the computation of the diagonal entries of $\tqH^H\tqH$ (required $NK$ multiplications), the computation of the matrix product of $ (\tqH\tqH^H+ \gamma^t\qI )^{-1}$ and $\tqH\tqH^H$ (required $K^3$ multiplications), the computation of the diagonal entries of $\tqH^H (\tqH\tqH^H+ \gamma^t\qI )^{-1} \tqH\tqH^H \tqH $ (required $N(K^2+K)$ multiplications), the computation of a scalar multiplication by $1/\gamma^{t}$ (required $N$ multiplications), and the computation of the inverse of the diagonal matrix (required $N$ multiplications). The total cost of this step is $ 2NK^2+\frac{4}{3}K^3+2NK+2N$.
Similarly, the $x$-update in line 6 requires $ 2NK+N$ multiplications. The
cost of projection $\Pi_{\calX}$ in the $x$-update is negligible. Line 7 involves $NK + K$ multiplications for computing $\|\qs-\tqH\qx^{t+1} \|^2_2$. The cost of the damping updates in lines 8 and 9 is negligible because multiplication by constant damping factor $\alpha$ can be implemented using a sequence of shifts and additions or subtractions.
Therefore, Algorithm 1 requires a total of $ 2NK^2+\frac{4}{3}K^3+5NK+3N+K$ multiplications for $t = 1$.
Given that $\tqH\tqH^H$ does not change in each iteration, we can cache the result to perform the subsequent iterations efficiently. Accordingly, Algorithm 1 requires a total of $NK^2+\frac{4}{3}K^3+5NK+3N+K$ multiplications for each iteration when $t \geq 2$. The total number of multiplications for
Algorithm 1 is $T(NK^2+\frac{4}{3}K^3+5NK+3N+K)+NK^2$, where $T$ denotes the number of iterations required to reach a stopping criterion. We summarize the number of multiplications of IDE in Table \ref{Table 1}.

Then, we analyze the complexity of IDE2 (Algorithm 2). The analysis of IDE2 is nearly similar to that of IDE, except for the fact that IDE2 does not require matrix inversion. In line 4 of Algorithm 2, we only form the reciprocal values of the diagonal entries of $\tqH^H\tqH$ and cache this result for the subsequent $x$-update, which only costs $NK+N$ multiplications. Line 5 requires $NK+N$ and $2NK+N$ multiplications to compute $\qW_{\rm u}\qs$ and $\qW_{\rm u}\tqH\qx^t$, respectively. Similarly, the costs for projection $\Pi_{\calX}$ and damping update are negligible. Therefore,
Algorithm 2 requires a total of $4NK+3N$ multiplications for $t = 1$. For $t \geq 2$, lines 4 and 5 only update $\qW_{\rm u}\tqH\qx_{d}^t$ because the other parts do not change. Accordingly, Algorithm 2 requires a total
of $2NK+N$ multiplications for each iteration when  $t \geq 2$. The overall complexity required by IDE2 is $T(2NK+N)+2NK+2N$ when performing $T$ iterations.

Following the analysis framework mentioned previously, we analyze the complexity of SQUID, C1PO, and TB-CEP and summarize the corresponding results in Table I. Notice that in the complexity analysis, we neglect the cost for calculating precoding factor $\beta$ because it is only updated every few iterations. In contrast to the other schemes, TB-CEP does not operate in an iterative manner. TB-CEP searches the precoding by using a trellis structure similar to that used by the Viterbi algorithm in decoding or channel equalization.
In each step, TB-CEP retains one path from $M^L$ possible combinations of the trellis states. The number of trellis states $M^L$ serves as a trade-off between complexity and performance.

We provide the total number of multiplications in Table II given specific values for system configurations to thoroughly understand the complexity of the mentioned schemes. We fix the number of users to $K = 16$ and show the results for two settings: i) $N = 64$ and ii) $N = 128$. The complexity of SQUID, C1PO, and the proposed algorithms depends on the number of iterations $T$.
For SQUID and C1PO, we set $T = 100$ and $24$, respectively, following the suggestions of their original proponents \cite{SQUID,TB_CEP}. For IDE and IDE2, we set $T = 100$, although good convergence is observed after approximately $50$ iterations.
Table II indicates that TB-CEP always exhibits relatively higher complexity than the others algorithms. {C1PO exhibits the better computational efficiency than the other algorithms for the  large MIMO system}.
IDE requires a slightly higher complexity than SQUID and C1PO but generally achieves excellent error-rate performance (to be shown subsequently in the simulations). {In the simulations, we show that IDE2 can achieve the better error-rate performance} than SQUID, C1PO, and TB-CEP. In addition, the complexity of IDE2 increases linearly {with the number of UE size $K$}. Consequently, IDE2 exhibits the best trade-offs between complexity and error-rate performance among all the compared algorithms.

\section*{ IV. Simulation Results and discussion}

\begin{figure}
    \centering
    \resizebox{3.5in}{!}{%
    \includegraphics*{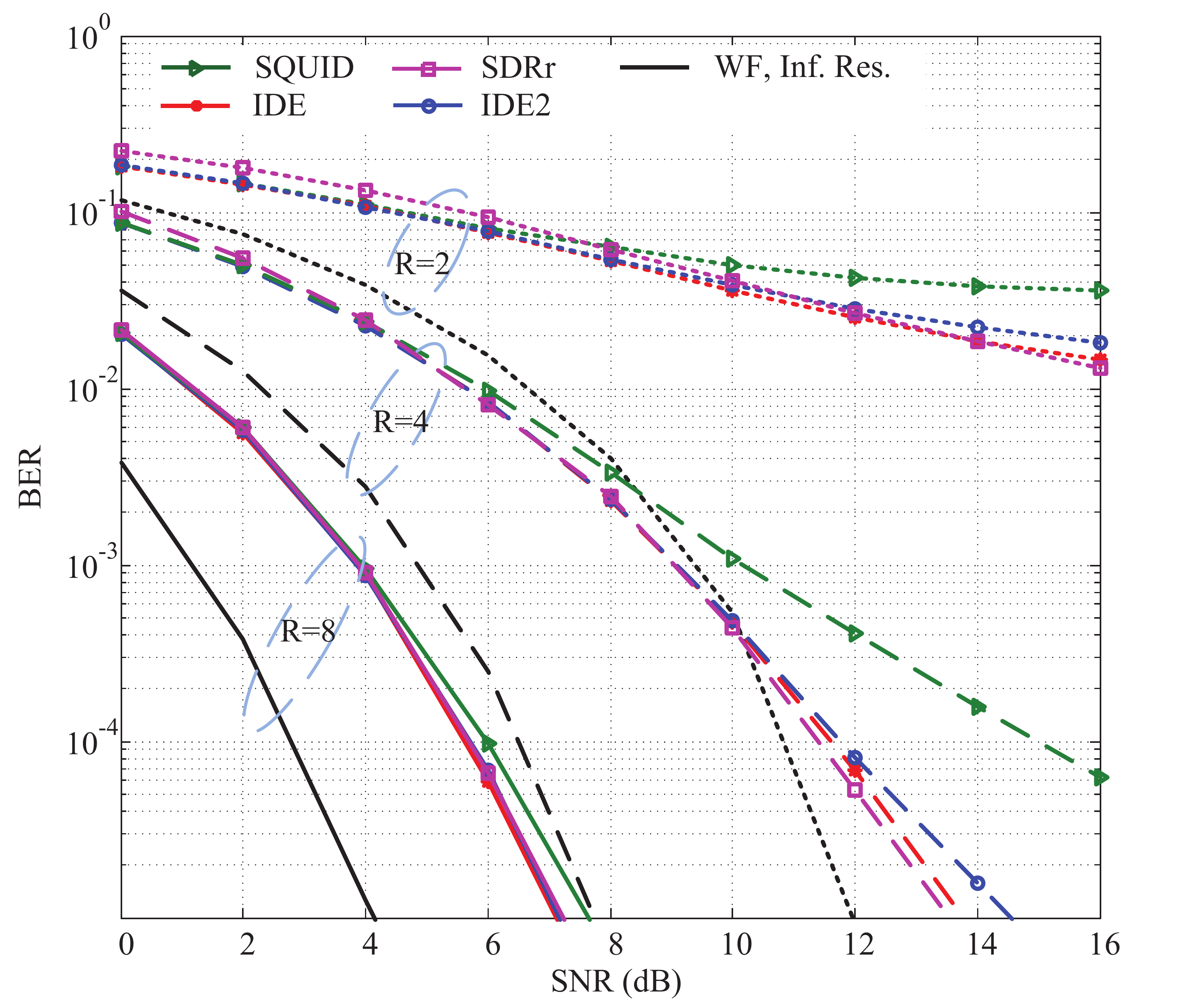} }%
    \caption{BER with QPSK signaling as a function of the SNR for different $1$-bit DAC precoders with adaptive $\beta$.
    The number of users is fixed to $K = 16$, and the total BS antenna number is scaled
    $N = R \times K$ from 32 to 64 and 128 by selecting $R=2,4,8$. \label{S_1}}
\end{figure}
\begin{figure}
    \centering
    \resizebox{3.5in}{!}{%
    \includegraphics*{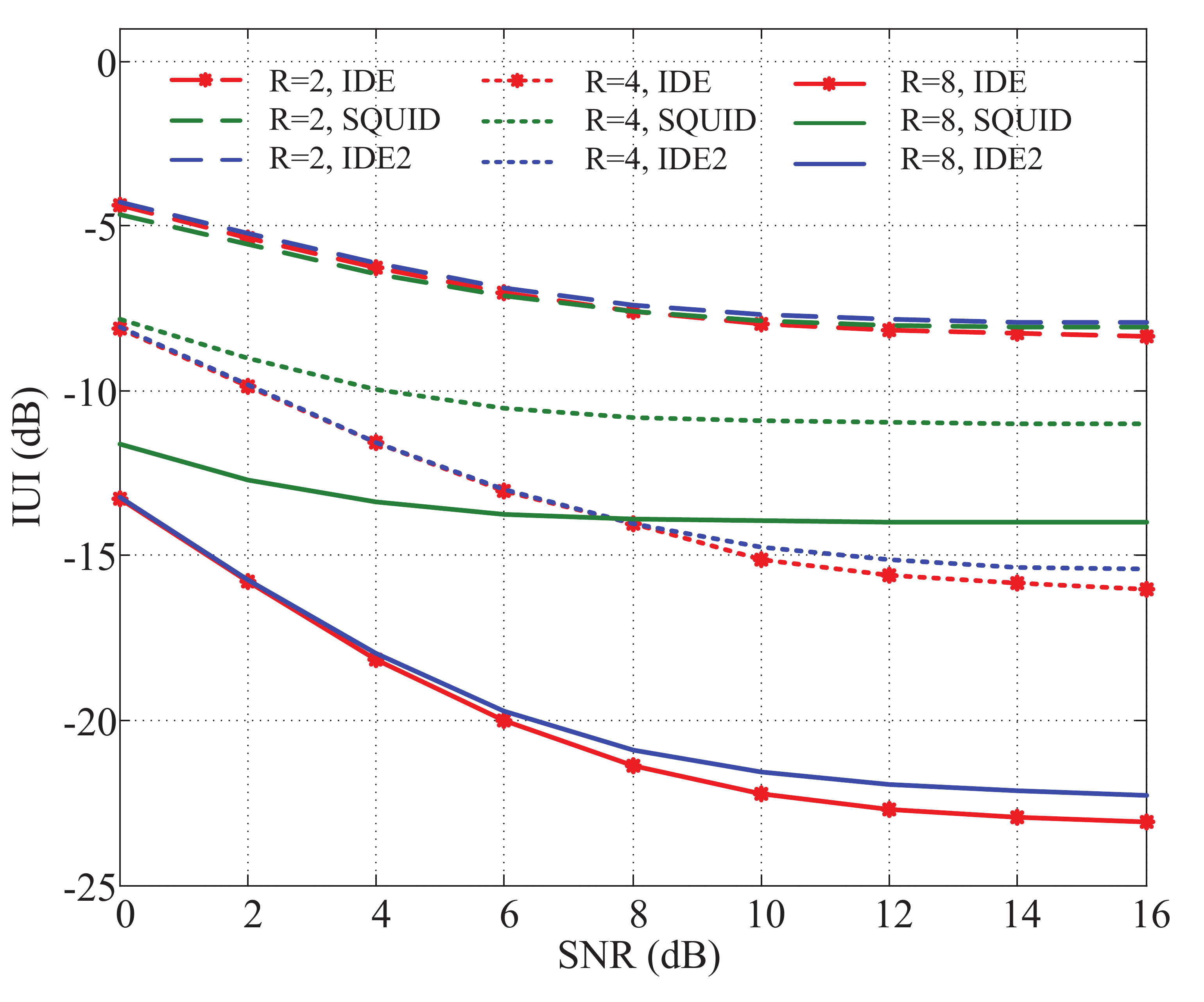} }%
    \caption{IUI versus SNR for different $1$-bit DAC precoders with adaptive $\beta$.\label{S_3}}
\end{figure}
\begin{figure}
    \centering
    \resizebox{3.5in}{!}{%
    \includegraphics*{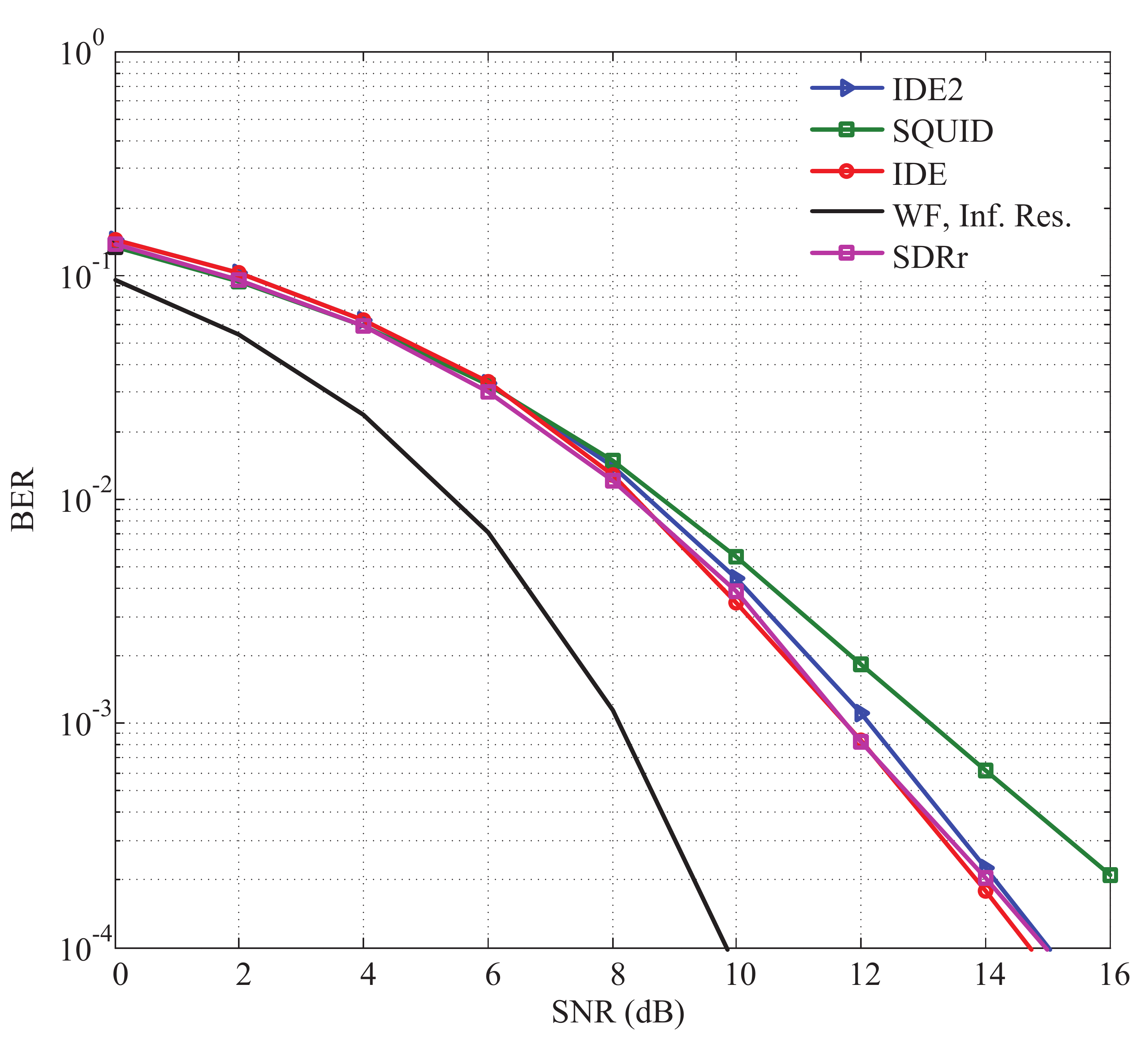} }%
    \caption{BER with 16-QAM signaling as a function of the SNR for different 1-bit DAC precoders with adaptive $\beta$; $K = 16$, $N=128$, and $R=8$.\label{S_9}}
\end{figure}

We conduct simulations to evaluate the performance of the proposed methods in terms of average IUI and BER. We consider a setting in which each antenna at the transmitter is equipped with a low-resolution DAC or PS. The 1-bit DAC is the extreme case of this setting. If no specification is provided, then the message symbol $\qs$ intended for each user is QPSK, and precoding factor $\beta$ is fixed to $1$.
We fix the number of users to $K = 16$ and scale the total BS antenna number
$N = R \times K$ from 32 to 64 and 128 by selecting $R=2,4,8$. We refer to $R$ as the system load factor.
The channel responses between the BS and each user follow a circular Gaussian distribution $\mathcal{CN}(0,1)$ in an i.i.d. manner.
Performances is evaluated via 30,000 channel realizations.
{Following \cite{R1}, we assume that factor $\beta$ is perfectly known for UEs because it can be estimated through pilot transmissions or blind estimation techniques.}
For IDE and IDE2, we set $T=100$, although good convergence is observed after approximately $50$ iterations.
Notably, the update of the precoding factor $\beta^{t}$ depends on $\qx^{t}$ as expressed in \eqref{eq:alpha}.
We begin with a fixed $\beta$ and update $\beta$ every $10$ iterations of $\qx^{t}$ to ensure that $\qx^{t}$ has converged to a good state.\footnote{This update setting is based on experience. Decreasing the iteration numbers of $\qx^{t}$ for each update of $\beta$ can facilitate escape for shallow local minima. However, fluctuations may occur. We also examine the damped update mechanism for $\beta$. This mechanism, however, degrades performance.}

\subsection*{A. 1-bit Precoding}

In Fig. \ref{S_1}, we compare the BER and SNR of our proposed algorithms and state-of-the-art methods, such as SDRr and SQUID \cite{SQUID}. As reported in \cite{SQUID}, the performance of SDRr is close to that observed in exhaustive search. Given that exhaustive search is impossible, we regard SDRr as a performance benchmark in this experiment. However, the computational complexity of SDRr is still high with the increase in the number of BS antenna $N$. In \cite{SQUID} and \cite{R2}, the authors proposed low-complexity versions called SQUID and C1PO, respectively. The performance of SQUID and C1PO is comparable. Thus, we only consider SQUID in the comparison.
In this figure, we use the adaptive precoding factor $\beta$ for all of the algorithms (including SDRr and SQUID). For comparison, we also report the performance of the WF precoder for the infinite-resolution case.

We observe that the BER performance of IDE is comparable to that of the benchmark SDRr. IDE has a low computational complexity.
The proposed algorithms (i.e., IDE and IDE2) present significant advantages in terms of BER compared with SQUID, particularly when $R$ is small (e.g., $R=2,4$). The performance of IDE2 is comparable to that of IDE, and IDE2 has a significantly lower computational complexity than IDE. When $R = 8$, the gap between the performance of the proposed algorithms and that of SQUID becomes negligible.
However, this result does not imply that their behaviors are similar.
We also show the corresponding IUIs for the three precoding algorithms in Fig. \ref{S_3} to thoroughly understand their differences.
When $R=8$, the proposed algorithms provide a significantly lower IUI than SQUID. The IUI of SQUID is saturated at -14\,dB. The error floor for IUI degenerates the BER performance when a high-modulation symbol (e.g., 16-QAM) is used because such a high modulation has to work in the high-SNR regime. In Fig.~\ref{S_9}, we compare the BER with the 16-QAM symbol and 1-bit DACs of our proposed algorithms and other algorithms under $R=8$ to justify this argument. Notably, the gaps between the performance of SQUID and our proposed algorithms increase with the increase in SNR because the IUI of SQUID saturates in the low-SNR regime (Fig. \ref{S_3}). In addition, the BERs of our proposed algorithms are similar to SDRr.

Another interesting observation from Fig. \ref{S_3} is that the IUIs of the three algorithms are similar when $R =2$. However, IDE and IDE2 offer significant advantages in terms of BER compared with SQUID (Fig. \ref{S_1}). The reason is that as $R=2$, their precoding factors $\beta$ are different and thus lead to different BERs. The precoding factor serves as a trade-off between IUI and noise enhancement. Therefore, this result indicates that IDE and IDE2 provide better trade-off in this regard than SQUID.

\subsection*{B. Robustness to Channel Estimation Errors}

\begin{figure}
\centering
 \resizebox{3.5in}{!}{ %
\includegraphics{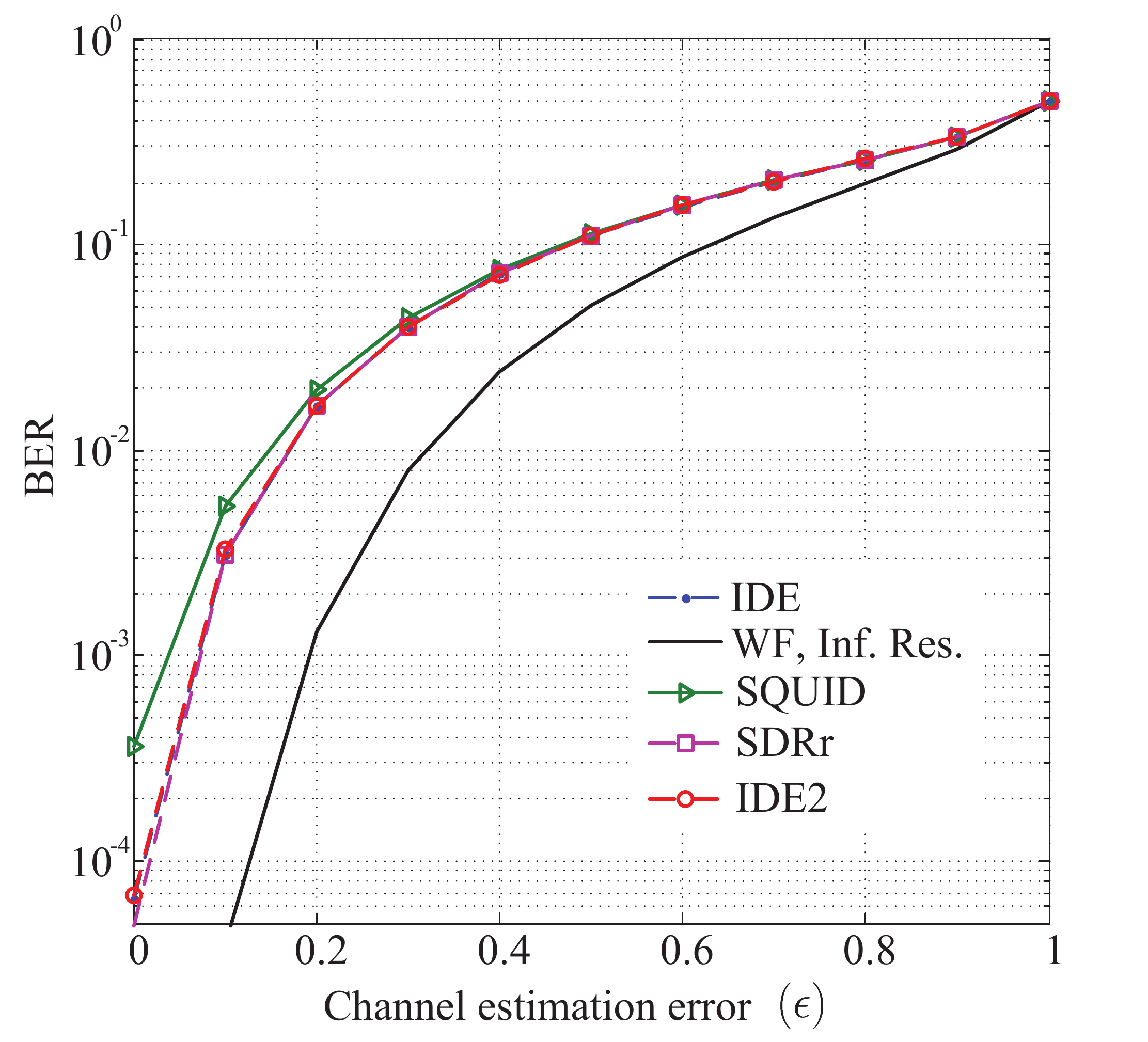} } %
 \caption{BER with QPSK signaling for channel estimation error $\epsilon$ under 1-bit DACs with SNR=$12$ dB and $(N,K,T,\beta) = (64,16,100,1)$.}\label{F-H-error}
\end{figure}

In this section, we present our investigation of the robustness of our proposed algorithms to channel estimation error. Following \cite{SQUID},
we assume that the BS acquires a noisy version of channel state information (CSI)
\begin{equation}
\hat{\qH} = \sqrt{1-\epsilon}\qH + \sqrt{\epsilon}\qE,
\end{equation}
where $\epsilon\in [0,1]$ is used to control the channel estimation error, and $\qE$ has $\mathcal{CN}(0,1)$ entries.
The value of $\epsilon=0$, $\epsilon \in (0,1)$, and $\epsilon=1$ correspond to cases with perfect CSI, partial CSI, or no CSI, respectively.
Fig. \ref{F-H-error} shows the BER with QPSK signaling as a function of the channel estimation error $\epsilon$ for SNR=$12$ dB and $(N,K,T,\beta) =(64,16,100,1)$.
Under imperfect CSI, the proposed IDE and IDE2 still outperform SQUID and are comparable with SDRr. This result is similar as that obtained under perfect CSI.

\subsection*{ C. Low-resolution PSs}

\begin{table}
\centering
\caption{Average running times in seconds}
\label{table1}
\begin{tabular}{|c|c|c|c|}
\hline
 Memory length & TB-CEP & IDE           & IDE2 \\ \hline
$L=3$                 & 0.0753     & \multirow{6}{*}{0.027 }          &\multirow{6}{*}{0.025 }                  \\ \cline{1-2}
$L=4$                 & 0.2538     &                            & \\ \cline{1-2}
$L=5$                 & 0.7546        &                         & \\ \cline{1-2}
$L=6$                 & 2.0994        &                          & \\ \cline{1-2}
$L=7$                 & 5.1339        &                          & \\ \cline{1-2}
$L=8$                 & 7.4925       &                           & \\ \hline
\end{tabular}
\end{table}

\begin{figure}
    \centering
    \resizebox{3.45in}{!}{%
    \includegraphics*{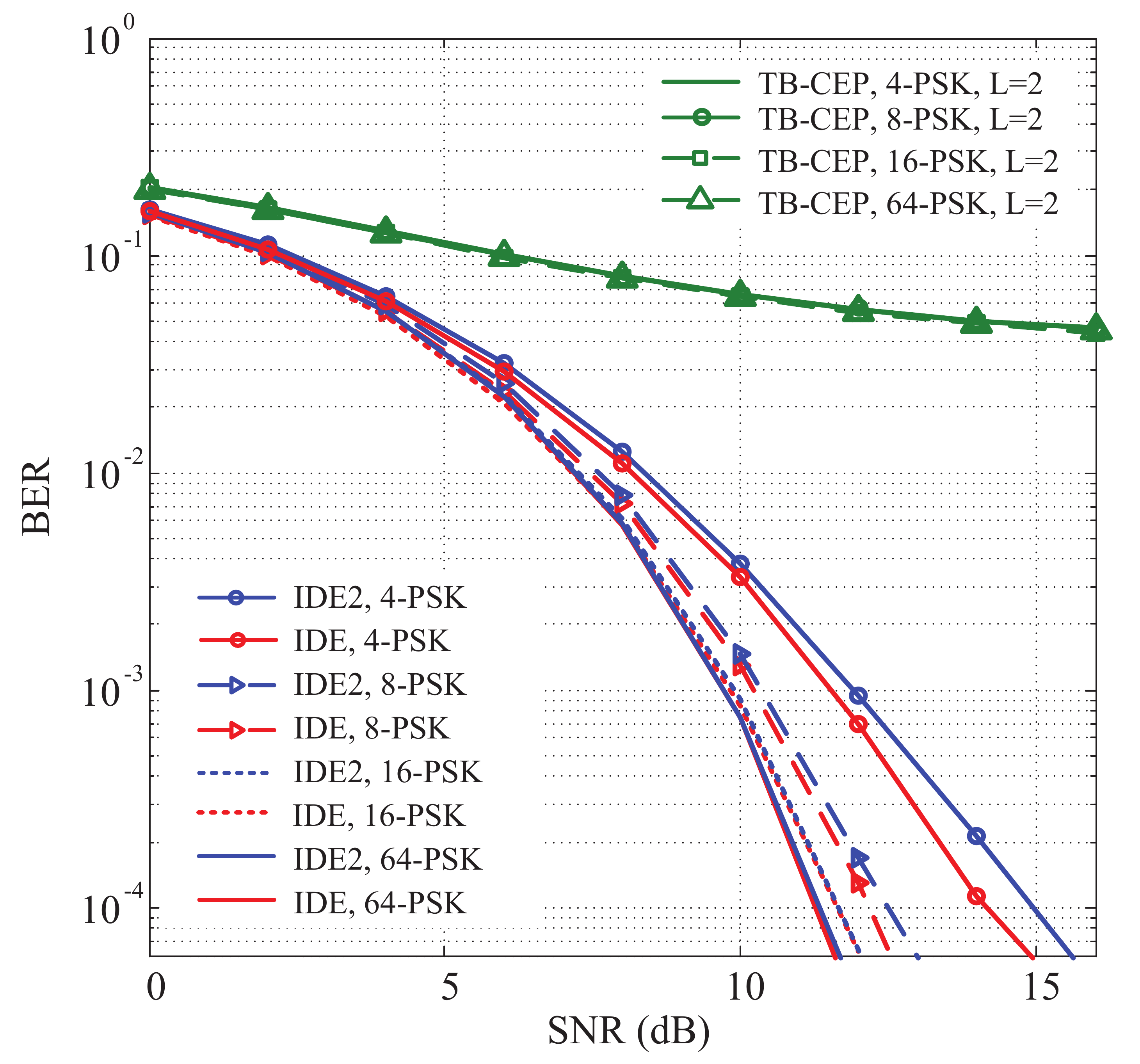} }%
    \caption{BER with QPSK signaling as a function of the SNR for different precoders with fixed $\beta$ under different PSK outputs; $K = 16$, $N=64$, and $R=4$.\label{S_4}}
\end{figure}

We examine our algorithms under a more general setting with finite-resolution PSs, that is, the possible values of the transmitting antenna outputs are from a set of $M$-PSK constellation points.
In this setting, we compare the proposed precoding algorithms with TB-CEP \cite{TB_CEP} rather than SQUID because SQUID cannot work in cases with general $M$-PSK outputs.
As previously analyzed in Section III.D, TB-CEP searches the precoding by using a trellis
structure, and the number of trellis states $M^L$ in TB-CEP serves as a trade-off between complexity and performance.
Following the commonly used setting in \cite{TB_CEP}, we take $L=3$.
In Fig. \ref{S_4}, we compare the BER of our proposed algorithms with those of TB-CEP \cite{TB_CEP} under different PSK outputs. Our proposed algorithms outperform TB-CEP. TB-CEP can achieve the same performance as our proposed algorithms when $L$ increases to a high level. However, the computational complexity of TB-CEP is too high to be practical. In addition, Fig. \ref{S_4} shows that IDE2 can be close to IDE as the number of resolution levels, $M$, increases.

\begin{figure}
    \centering
    \resizebox{3.45in}{!}{%
    \includegraphics*{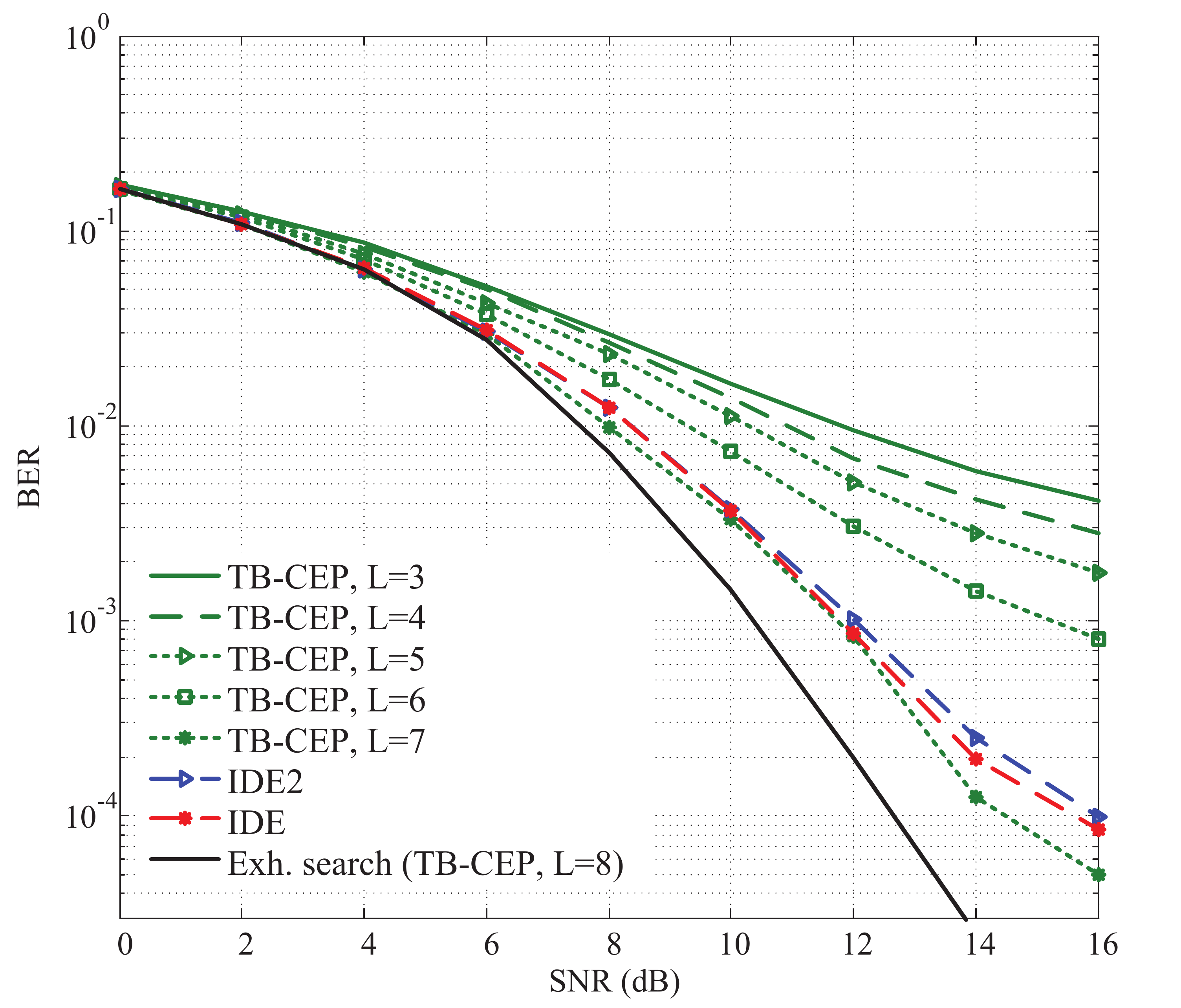} }%
    \caption{BER with QPSK signaling as a function of the SNR for different precoders with the four-phase state of the PSs; $K = 2$, $N=8$, and $R=4$. \label{S_5}}
\end{figure}

We consider a small MU-MIMO system with eight BS antennas and two users, that is, $N=8$ and $K = 2$, to determine the gaps between the optimal precoder (exhaustive search) and the considered algorithms.
The optimal precoding can be obtained by exhaustive search or TB-CEP with $L=8$. Fig. \ref{S_5} compares the BERs of IDE, IDE2, TB-CEP, and the optimal precoder when the four-phase state of the PSs is employed.
We determine that the gap between the performance of the optimal precoder and the performance of IDE (or IDE2) is small, namely, approximately 2\,dB for a target BER of $10^{-3}$.
When $L = 7$, TB-CEP presents a comparable performance to that of the proposed algorithms. However, the computational complexity of TB-CEP is too high to be practical when the number of BS antennas is large.
We summarize the average running times\footnote{The simulations are performed with MATLAB v8.6.0 (R2015b) on a 64-bit Windows 7 PC equipped with a 3.4-GHz Intel Core i7-3370 CPU and 4 GB of memory.} (in seconds) of the algorithms in Table \ref{table1}. IDE and IDE2 provide significant advantages in terms of complexity compared with TB-CEP.

\begin{figure*}[!ht]
    \centering
    \resizebox{5.3in}{!}{%
    \begin{tabular}{ccc}
 \includegraphics*{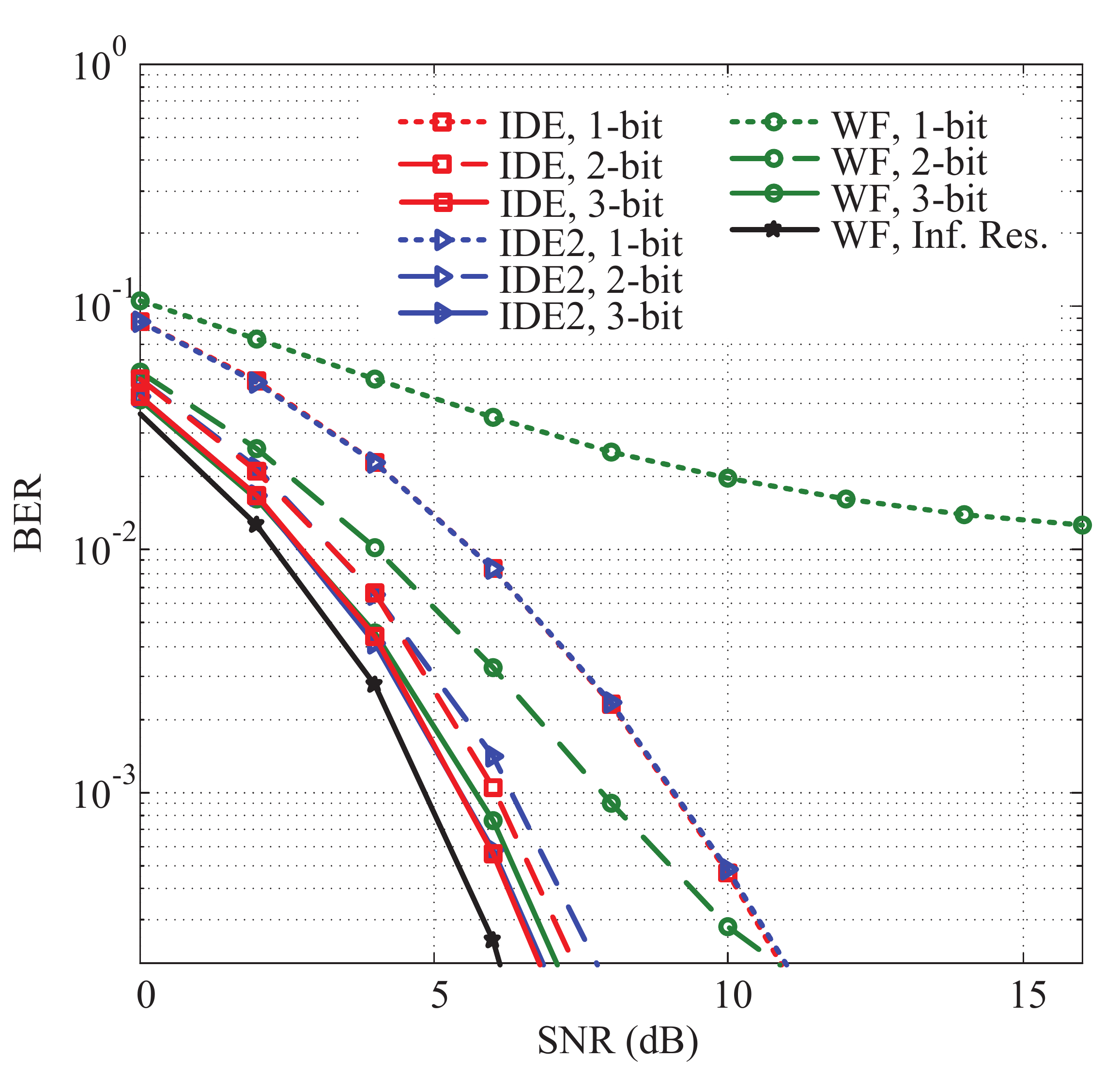}              & \includegraphics*{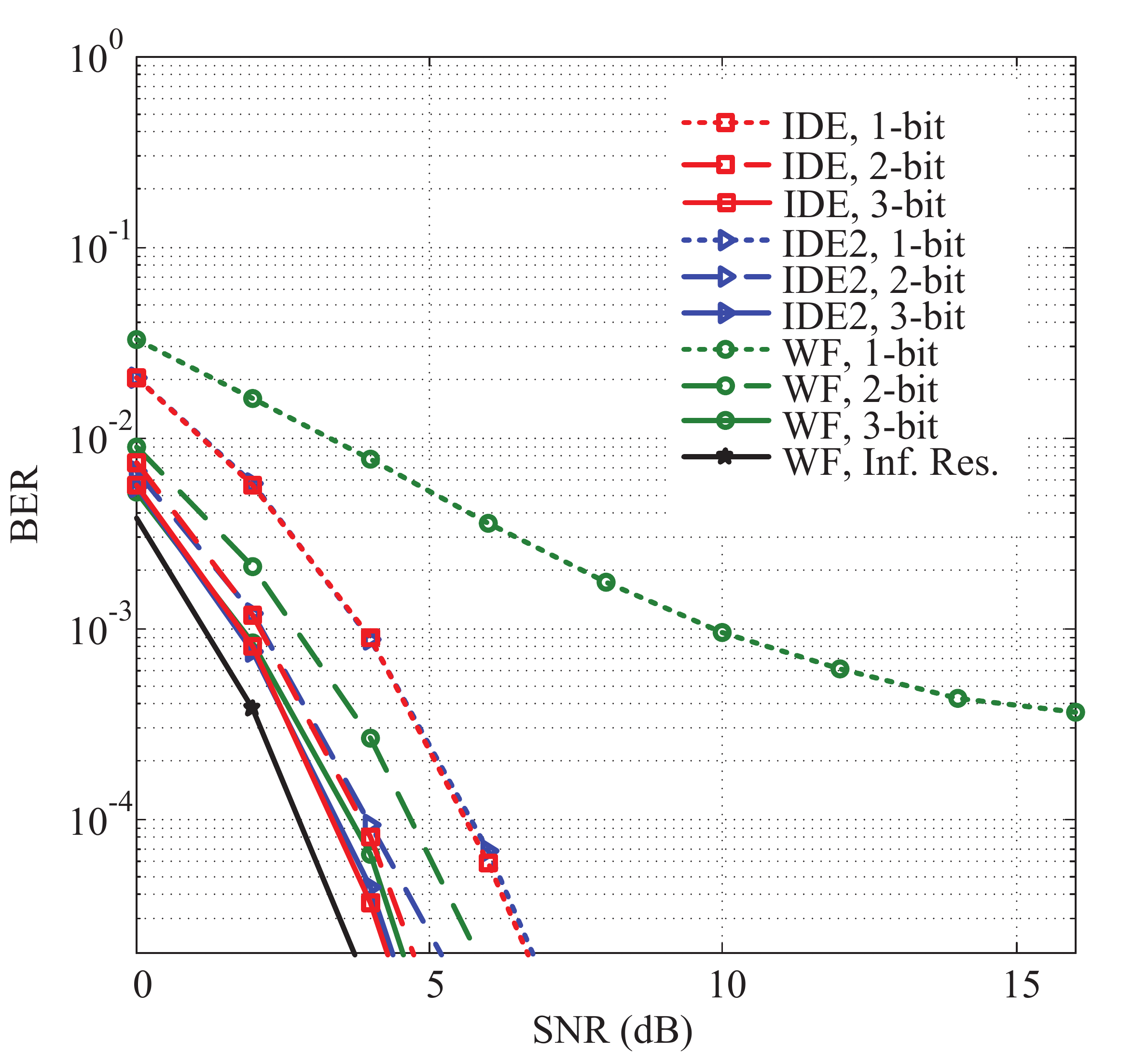}     \\
  \Huge (a) QPSK, $K = 16$, $N=64$, and $R=4$& \Huge (b) QPSK, $K = 16$, $N=128$, and $R=8$     \\
   \includegraphics*{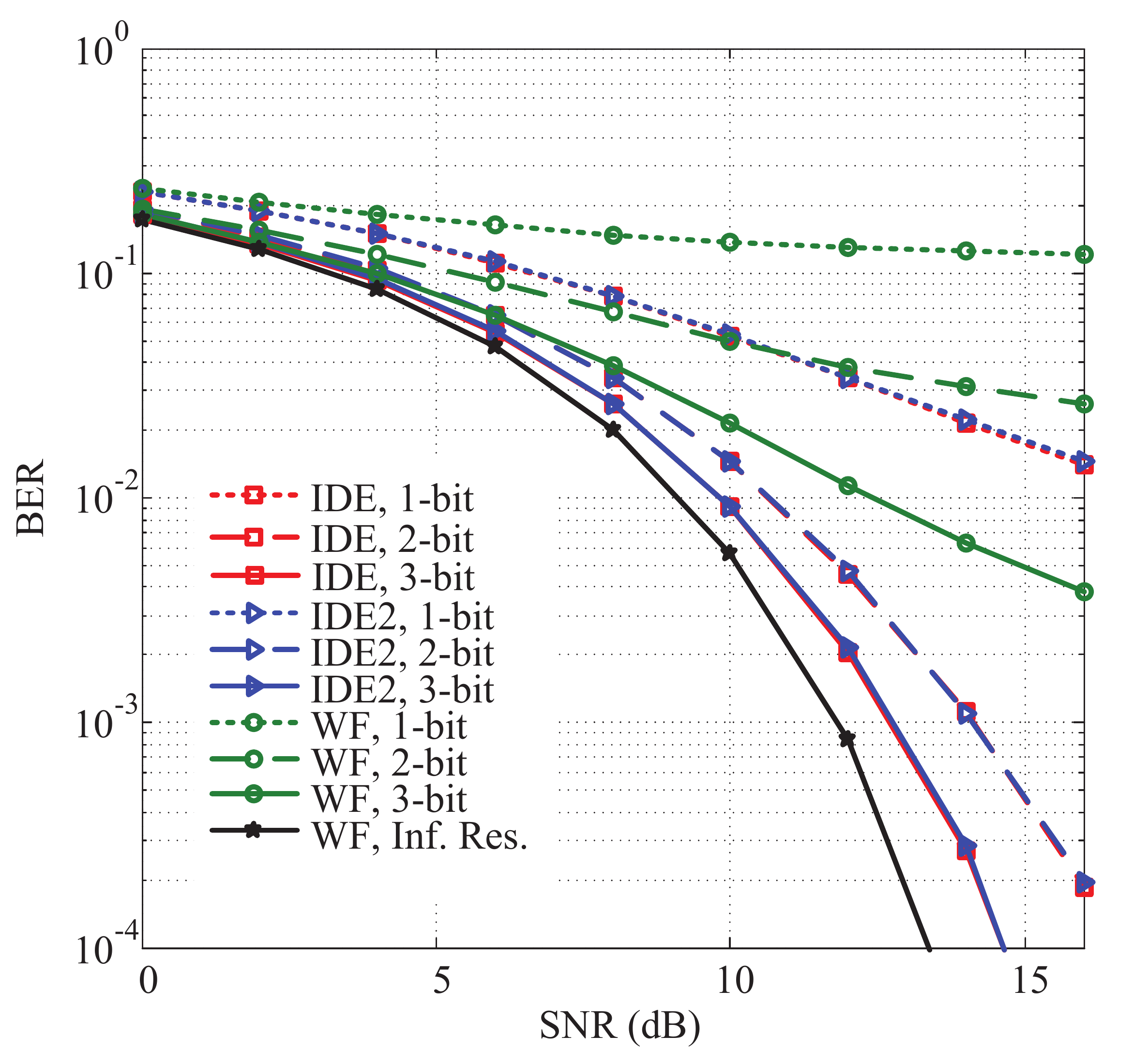}       & \includegraphics*{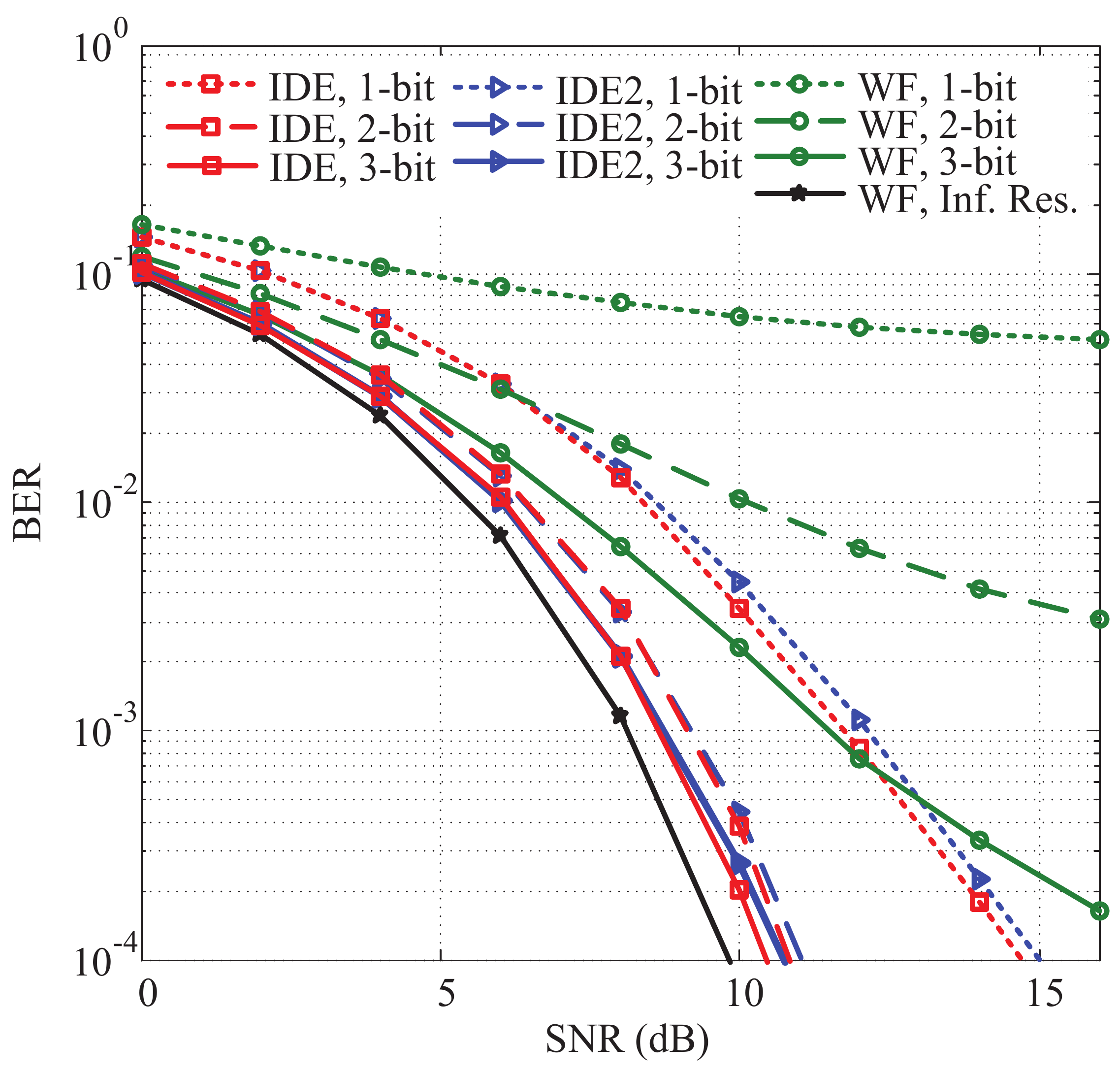} \\
  \Huge (c) 16-QAM, $K = 16$, $N=64$, and $R=4$ & \Huge (d) 16-QAM, $K = 16$, $N=128$, and $R=8$\\
    \includegraphics*{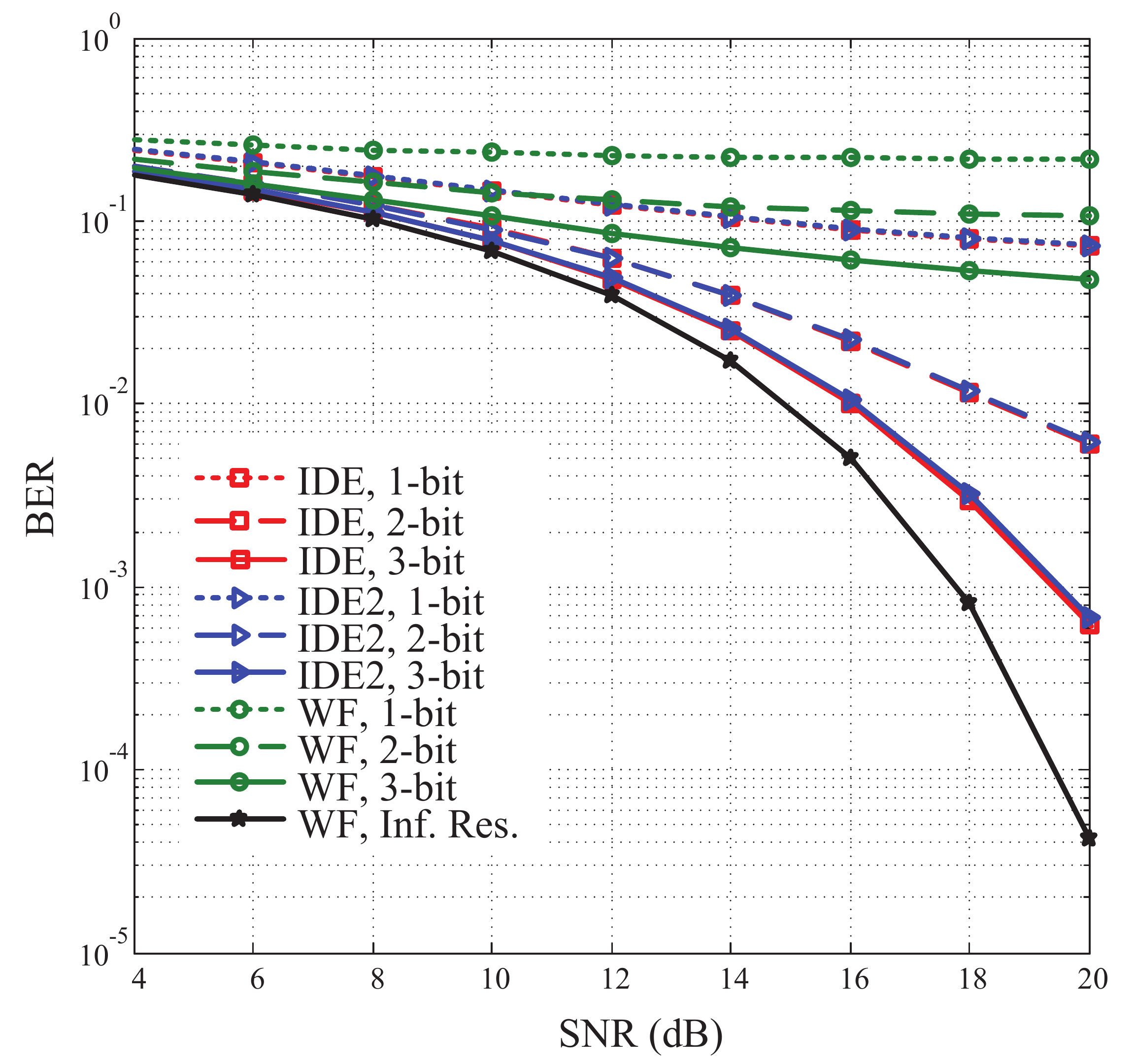}      &  \includegraphics*{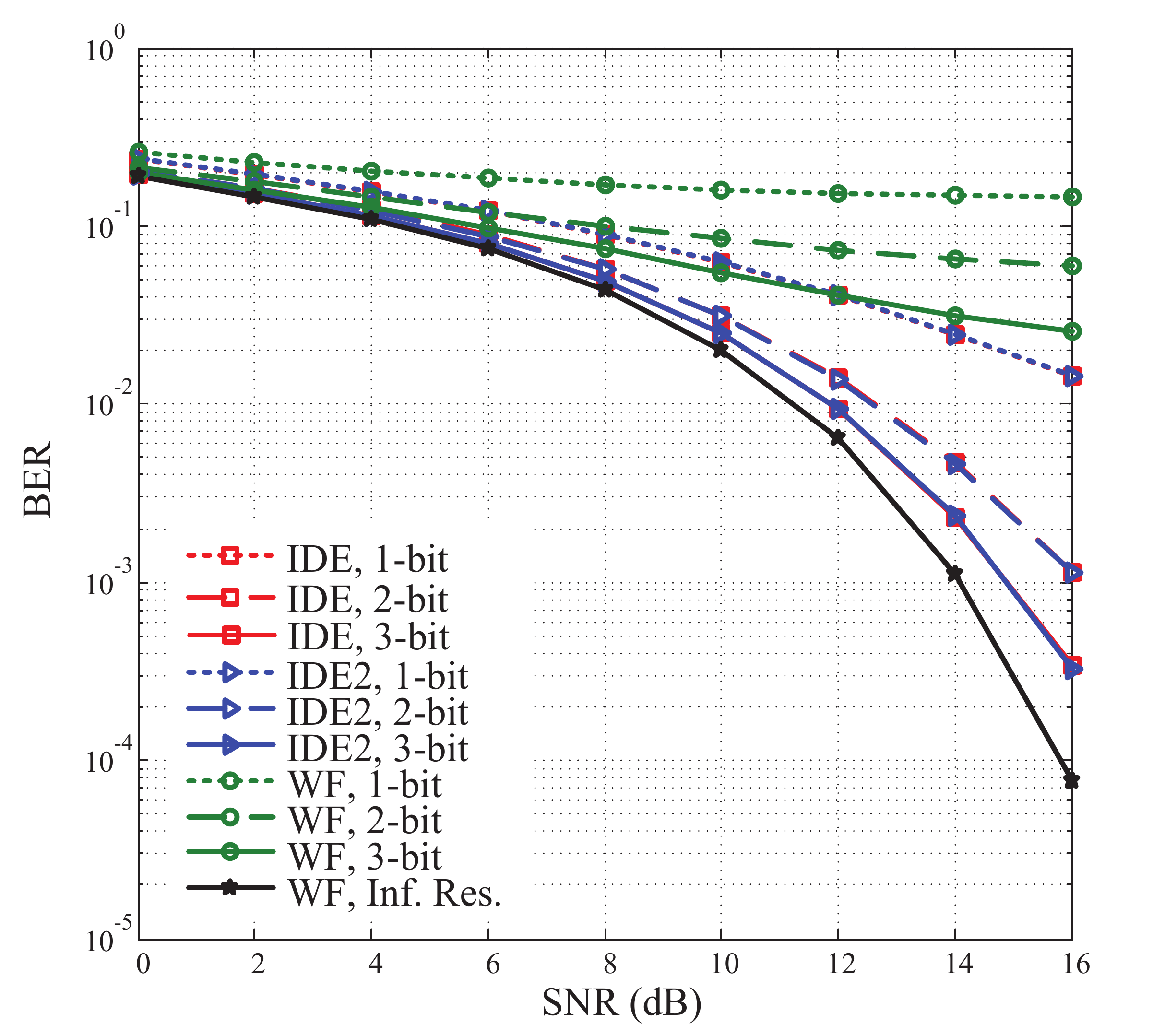}          \\
  \Huge (e) 64-QAM, $K = 16$, $N=64$, and $R=4$ & \Huge (f) 64-QAM, $K = 16$, $N=128$, and $R=8$\\
  \end{tabular}
    }%
    \caption{BER as a function of the SNR for different DAC resolution precoders with adaptive $\beta$.\label{S_7_1}}
\end{figure*}

\subsection*{D. Low-resolution Precoding}

In the previous experiments, the algorithms are evaluated by using either 1-bit DACs or PSs at the transmitters.
Both of them are CE precoding. Then, we shift our attention to multi-bit precoding for low-resolution DACs, in which the precoding has multiple amplitude levels.
When low-resolution DACs are employed, \cite{SQUID} suggests quantizing the values of WF precoding to a finite set directly. In Fig. \ref{S_7_1}, we compare the BERs of the quantized WF precoding with those of IDE and IDE2 under different DAC resolution precoders, modulation symbols, and load factors.

First, we focus on the case with 4-QAM signaling (Figs. \ref{S_7_1}(a) and \ref{S_7_1}(b)).
When $R=4$, the 1-bit WF precoder cannot achieve BER below $10^{-3}$ and has significant gaps compared with
IDE and IDE2. For 2-bit DACs, IDE and IDE2 gain $2$\,dB compared with the 2-bit WF precoder for a target BER of $10^{-3}$. In the case with $3$-bit DACs, all algorithms perform similarly, but IDE and IDE2 exhibit the better performance. In this case, all algorithms are close to the limit performance of the infinite-resolution WF precoder.
Similar characteristics are observed when $R=8$.
Second, we focus on cases with 16-QAM and 64-QAM signaling (Figs. \ref{S_7_1}(c) to \ref{S_7_1}(f)). IDE and IDE2 have a significant performance gain over the quantized WF precoder in all quantization levels. For 3-bit DACs, IDE and IDE2 are close to the limit performance of infinite-resolution WF precoding, whereas the 3-bit WF precoder is still far from the limit performance. Moreover, the performance of IDE2 is compared with the performance of IDE while exhibiting significantly lower computational complexity.

The discussions in the previous subsections focus on the 1-bit precoder, low-resolution PSs, and multi-bit precoder. In prior state-of-the-art methos, the precoding techniques for different settings are completely different. Notably, our algorithms are not only universal but can also perform well in various finite-alphabet precoders with general QAM signaling.

\begin{figure}
    \centering
    \resizebox{3.45in}{!}{%
    \includegraphics*{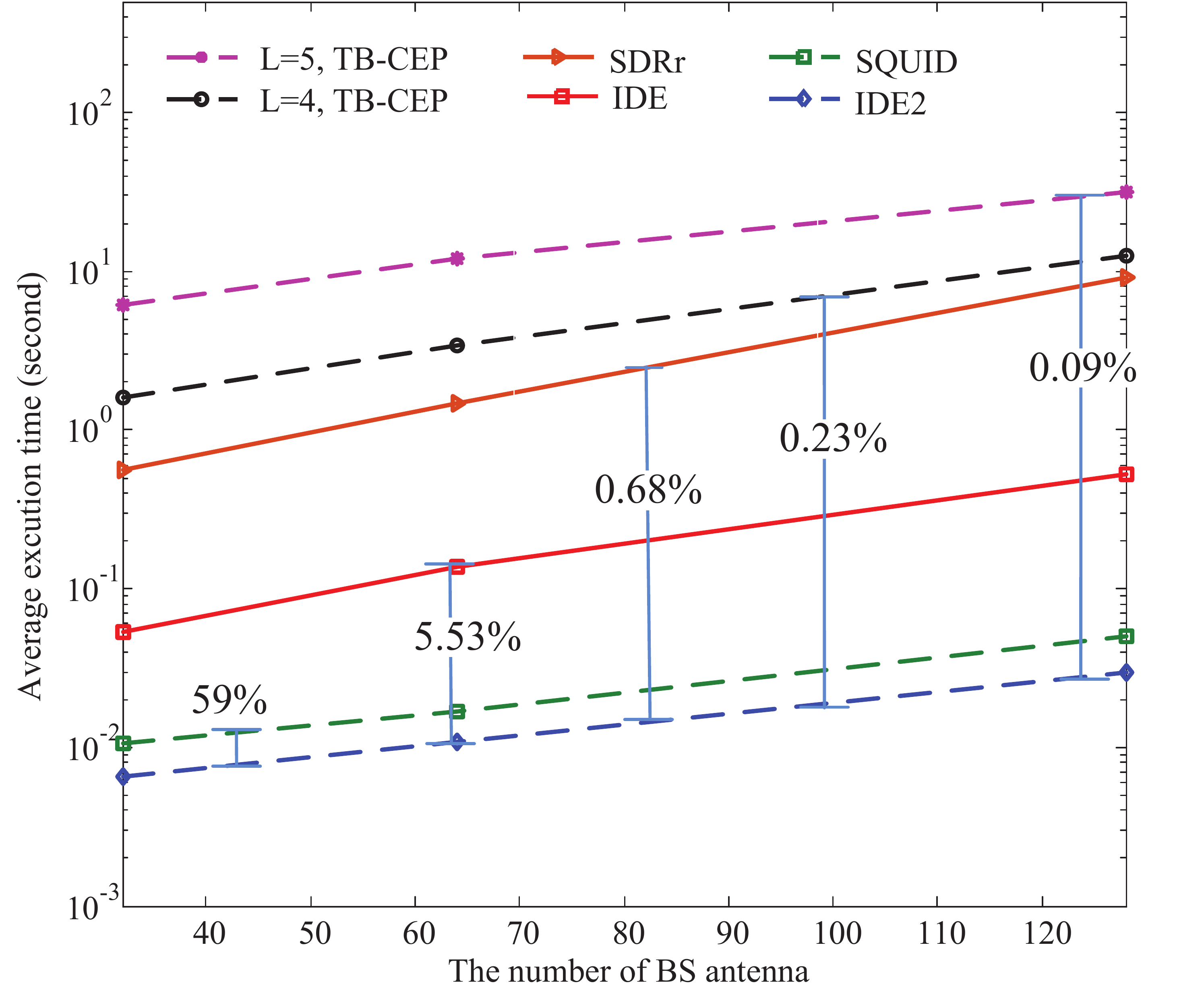} }%
    \caption{Average execution time versus the number of BS antennas for all of the concerned precoders. We consider PSs with four-phase state and $T =100$. \label{S_6}}
\end{figure}

\subsection*{E. Complexity and Convergence Rate}
Fig. \ref{S_6} compares the average running times versus $N$ for all of the concerned algorithms. We consider PSs with four-phase state and $T =100$.
IDE2 exhibits the lowest timing complexity among the algorithms, followed by SQUID and IDE. The timing complexity of SDRr and TB-CEP increases significantly with the number of BS antennas.
After integrating all of the previous experiments, we conclude that IDE2 exhibits the best trade-off between performance and complexity among all the algorithms.

\begin{figure}
    \centering
    \resizebox{3.45in}{!}{%
    \includegraphics*{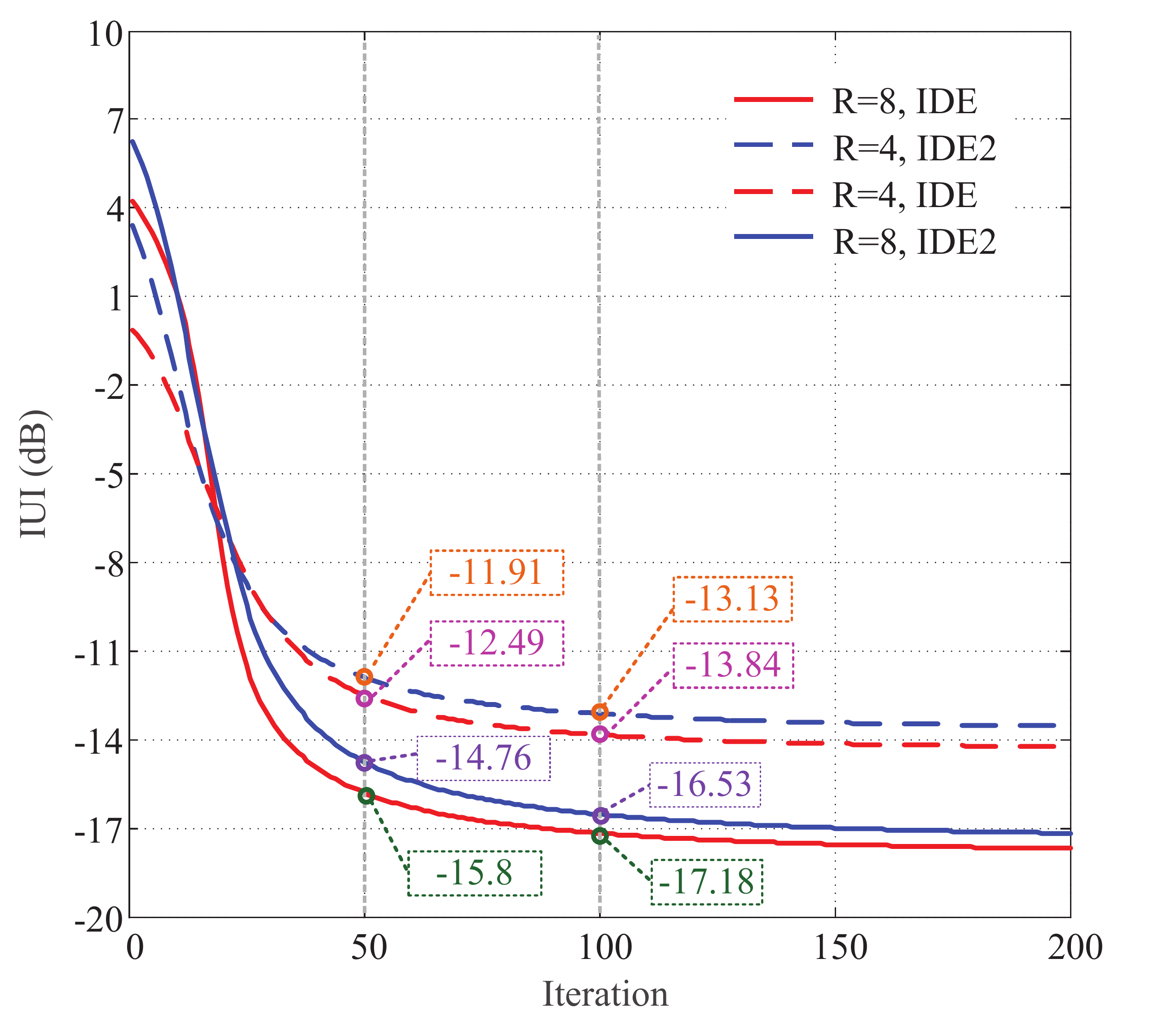} }%
    \caption{IUI versus iteration under 1-bit DACs with fixed $\beta$ for $R=4,8$ and {SNR $=0$ dB}.\label{S_8}}
\end{figure}

Timing complexity heavily depends on the number of iterations $T$.
As previously mentioned, IDE and IDE2 perform well after approximately 50 iterations.
Fig. \ref{S_8} shows the IUI versus iteration under 1-bit DACs with fixed $\beta$ for $R=4,8$ and {SNR $=0$\,dB} to justify this argument. We observe that IUI only decreases by approximately $1$ dB as the iterations increase from $50$ to $100$. The IUI at $T=50$ is sufficient to obtain good detection for $4$-QAM or $16$-QAM symbols.
In addition, $\qx$ does not obtain a good IUI result after the first few iterations. Therefore, as mentioned at the beginning of this section, we update $\beta$ after every $10$ iterations of $\qx^{t}$.

\section*{ V. Conclusion}

We propose novel precoding algorithms called IDE and IDE2 for a massive MU-MIMO system where each antenna at the BS is equipped with coarse-resolution DACs or PSs. The algorithms have a unified structure, such that they can be applied to various finite-alphabet input problems that aim to minimize IUI. Compared with state-of-the-art methods, the proposed precoding algorithms show significant advantages in terms of performance and computational complexity. In particular, IDE demonstrates excellent performance and is comparable with the benchmark in all test scenarios.
The performance of IDE2 is comparable with that of IDE, whereas IDE2 is simple and exhibits low per-iteration complexity.
From the perspective of complexity-performance trade-off, IDE2 is highly suitable for massive MU-MIMO systems. Future work can include channel estimation error (e.g., the variance of the error) into the optimization problem, and robust precoding algorithms for channel estimation error can be designed.

\section*{Appendix A: Approximately Unbiased Estimator}

This appendix aims to show that \eqref{eq:Proposed1-1} is an approximately unbiased estimator of $\qx$.
We begin by showing that (\ref{eq:bias_lmmse}) can be
interpreted as the optimal linear MMSE estimate of $\qx$ given
prior knowledge on (\ref{eq:priorOfx}). To this end, we introduce the following virtual channel model
\begin{equation} \label{eq:virtualCh}
    \qs = \qH\qx + \tilde{\qz},
\end{equation}
where $\tilde{z}_k$'s are i.i.d circularly-symmetric complex Gaussian with mean $0$ and variance $1$, that is, $\tilde{\qz} \sim \mathcal{CN}(0,\qI)$.
Suppose that we wish to construct an estimator for $\qx$ from $\qs$. We restrict our estimator to be in the form
\begin{equation}
    \hat{\qx} = \qW\qs +\qb,
\end{equation}
where matrix $\qW$ and vector $\qb$ are to be determined to minimize $\Ex\{ (\hat{\qx}-\qx)^H(\hat{\qx}-\qx)\}$. Notably, the expectation is taken over the joint density of $\qs$ and $\qx$.
The optimal $\qW$ and $\qb$ are expressed as follows \cite{Unbiased}:
\begin{equation}
\qW = \qC_{xs}\qC_{s}^{-1}, \quad
\qb = \bar{\qx}-\qW\bar{\qs},
\end{equation}
where $\bar{\qx} = \Ex\{ \qx \}$, $\bar{\qs} = \Ex\{ \qs \}$, $\qC_{xs}$ is the cross-covariance matrix between $\qx$ and $\qs$, and $\qC_{s}$ is the auto-covariance matrix of $\qs$.
Specifically, by substituting \eqref{eq:virtualCh}, we obtain
\begin{equation}
\begin{aligned}
\qC_{xs} &= \Ex\{(\qx -\bar{\qx})(\qs -\bar{\qs})^H \},\\
			 &=\Ex\{(\qx -\bar{\qx})(\qx -\bar{\qx})^H\tqH^H + (\qx -\bar{\qx})\tilde{\qz}^H \},\\
			 &=\qC_{x}\tqH^H,
\end{aligned}
\end{equation}
and
\begin{equation}
\begin{aligned}
\qC_{s} &=\Ex\{(\qs -\bar{\qs})(\qs -\bar{\qs})^H \}, \\
	      &=\Ex\left\{ \left(\tqH(\qx -\bar{\qx})+\tilde{\qz}\right)\left(\tqH(\qx -\bar{\qx})+\tilde{\qz}\right)^H \right\},\\
	      &=\tqH\qC_{x}\tqH^H +\qC_{\tilde{z}}.
\end{aligned}
\end{equation}

Using prior knowledge on $\qx$ in (\ref{eq:priorOfx}) and $\qC_{\tilde{z}} = \qI$, we obtain
\begin{subequations}
\begin{align}
\qW &= \qC_{xs} \qC_{s}^{-1}
	=\tqH^H(\tqH\tqH^H +\gamma\qI)^{-1} \notag \\
	& =(\tqH^H\tqH +\gamma\qI)^{-1}\tqH^H, \\
\qb & = \qx_{\rm d}^t - \qW \tqH\qx_{\rm d}^t.
\end{align}
\end{subequations}
The optimal linear MMSE estimate is
\begin{equation} \label{eq:lmmse_estimator}
\hat{\qx} = \bar{\qx} + \qW (\qs-\bar{\qs})
		  = \qx_{\rm d}^t + \qW(\qs-\tqH\qx_{\rm d}^t),
\end{equation}
which is completely identical to (\ref{eq:bias_lmmse}).

Taking the expected value on both sides of \eqref{eq:lmmse_estimator} produces
\begin{align}
\Ex\{ \hat{\qx} \} &= \Ex\{\qx_{\rm d}^t-\qW\tqH\qx_{\rm d}^t+\qW\qs\}, \notag \\
		&= (\qI-\qW \tqH) \qx_{\rm d}^t+\qW \tqH\Ex\{ \qx \} \neq \Ex\{ \qx \}. \label{eq:hatx2}
\end{align}
Clearly, ${\qx}$ is a biased estimator of $\qx$ as long as $\Ex\{ \qx \} \neq \qx_{\rm d}^t$. $\qx_{\rm d}^t$ is obtained from the previous iteration, which could not be equal to $\qx_{\rm d}^t$ before convergence.
We can remove the bias of $\hat{\qx}$ by setting $\gamma=0$. In this case, \eqref{eq:lmmse_estimator} becomes
\begin{equation}
\hat{\qx} = (\tqH^H\tqH)^{-1}\tqH^H\qs.
\end{equation}
Moreover, we obtain $\Ex\{ (\tqH^H\tqH)^{-1}\tqH^H\qs \} = \Ex\{ \qx \}$.
However, this approach degenerates the iteration algorithm because it causes the estimation step to completely ignore knowledge on the previous iteration result $\qx_{\rm d}^t$.

From the discussion above, we realize that factor $\gamma$ should be preserved.
However, as long as $\gamma$ appears, the form expressed in \eqref{eq:hatx2} always presents a biased estimate because the main diagonal elements of $\qW\tqH$ are smaller than $1$.
To \emph{approximately} remove the bias, we normalize the diagonal elements of $\qW\tqH$ as follows:
\begin{equation}\label{eq:normalize matrix}
\qD\qW\tqH = [\diag(\qW\tqH)]^{-1}\qW\tqH.
\end{equation}
By doing so, we obtain $\hat{\qx}  = \qx_{\rm d}^t + \qD\qW(\qs-\tqH\qx_{\rm d}^t)$ and
\begin{equation}
 \Ex\{\hat{\qx}\} =(\qI-\qF)\qx_{\rm d}^t+\qF\Ex \{\qx \},
\end{equation}
where $\qF = \qD\qW\tqH$.
Given that the diagonal elements of $\qF$ are $1$ and the off diagonal elements of $\qF$ are smaller than $1$, we obtain $(\qI-\qF)\qx_{\rm d}^t \approx \qzero$, $\qF\Ex\{\qx \} \approx \Ex \{\qx \}$, and $\Ex\{\hat{\qx}\} \approx \Ex \{\qx \} $. According to the central limit theorem, this approximation becomes good when $N$ becomes large. The new estimator is an approximately unbiased estimator.

\section*{Appendix B: Derivations of $\beta$}

When $\qx$ is given, the minimum problem (\ref{eq:unqMMSE1}) can be written in the form:
\begin{align}
\min_{\beta > 0}&\ \ \|\qs -\beta\qH\qx \|^2_2 +\beta^2 K \sigma^2.
\end{align}
We unfold the objective function as follows:
\begin{multline}
\left(\qs-\beta\qH\qx\right)^H \left(\qs-\beta\qH\qx\right) + \beta^2 K \sigma^2 \\
=\ \qs^H\qs-2\beta\rm{Re}{\left\{\qs^H\qH\qx\right\}}+\beta^2 \|\qH\qx\|_2^2 +\beta^2 K \sigma^2. \label{eq:diff}
\end{multline}
By equating the differential of (\ref{eq:diff}) with respect to $\beta$ to zero, we obtain
\begin{equation}
 -2\rm{Re}\left\{\qs^H\qH\qx\right\} + 2\beta \|\qH\qx\|_2^2+2\beta K\sigma^2 = 0.
\end{equation}
If $\|\qH\qx\|_2^2 + K\sigma^2 \neq 0$, then we obtain $\beta =\frac{\rm{Re}\left(\qs^H\qH\qx\right)}{ \|\qH\qx\|_2^2+K\sigma^2 }$.

{\renewcommand{\baselinestretch}{1.1}

}


\begin{thebibliography}{10}
\providecommand{\url}[1]{#1}
\csname url@samestyle\endcsname
\providecommand{\newblock}{\relax}
\providecommand{\bibinfo}[2]{#2}
\providecommand{\BIBentrySTDinterwordspacing}{\spaceskip=0pt\relax}
\providecommand{\BIBentryALTinterwordstretchfactor}{4}
\providecommand{\BIBentryALTinterwordspacing}{\spaceskip=\fontdimen2\font plus
\BIBentryALTinterwordstretchfactor\fontdimen3\font minus
  \fontdimen4\font\relax}
\providecommand{\BIBforeignlanguage}[2]{{%
\expandafter\ifx\csname l@#1\endcsname\relax
\typeout{** WARNING: IEEEtran.bst: No hyphenation pattern has been}%
\typeout{** loaded for the language `#1'. Using the pattern for}%
\typeout{** the default language instead.}%
\else
\language=\csname l@#1\endcsname
\fi
#2}}
\providecommand{\BIBdecl}{\relax}
\BIBdecl

\bibitem{centralized1}
T.~L. Marzetta, ``Noncooperative cellular wireless with unlimited numbers of
  base station antennas,'' \emph{IEEE Trans. Wireless Commun.}, vol.~9, no.~11,
  pp. 3590--3600, Nov. 2010.

\bibitem{centralized2}
J.~Hoydis, S.~ten Brink, and M.~Debbah, ``Massive {MIMO} in the {UL}/{DL} of
  cellular networks: How many antennas do we need?'' \emph{IEEE J. Sel. Areas
  Commun.}, vol.~31, no.~2, pp. 160--171, Feb. 2013.

\bibitem{distributed}
J.~Zhang, C.~K. Wen, S.~Jin, X.~Gao, and K.~K. Wong, ``On capacity of
  large-scale {MIMO} multiple access channels with distributed sets of
  correlated antennas,'' \emph{IEEE J. Sel. Areas Commun.}, vol.~31, no.~2, pp.
  133--148, Feb. 2013.

\bibitem{I4}
R.~H. Walden, ``Analog-to-digital converter survey and analysis,'' \emph{IEEE
  J. Sel. Areas Commun.}, vol.~17, no.~4, pp. 539--550, Apr. 1999.

\bibitem{DAC-power}
C.~Toumazou, G.~S. Moschytz, and B.~Gilbert, \emph{Trade-Offs in Analog Circuit
  Design: The Designer's Companion}.\hskip 1em plus 0.5em minus 0.4em\relax
  Kluwer Academic, 2002, pp. 592--594.

\bibitem{DAC-power2}
Y.~Li, B.~Bakkaloglu, and C.~Chakrabarti, ``A system level energy model and
  energy-quality evaluation for integrated transceiver front-ends,'' \emph{IEEE
  Trans. Very Large Scale Integr. Syst.}, vol.~15, no.~1, pp. 90--103, Jan
  2007.

\bibitem{Energy-Efficiency}
\BIBentryALTinterwordspacing
L.~N. Ribeiro, S.~Schwarz, M.~Rupp, and A.~L.~F. de~Almeida, ``{Energy
  efficiency of mm{Wave} massive {MIMO} precoding with low-resolution
  {DAC}s},'' 2017. [Online]. Available: \url{https://arxiv.org/abs/1709.05139}
\BIBentrySTDinterwordspacing

\bibitem{syn}
A.~Wadhwa and U.~Madhow, ``Blind phase/frequency synchronization with
  low-precision {ADC}: A bayesian approach,'' in \emph{Proc. 51st Allerton
  Conf. Communication, Control and Computing}, IL, USA, Oct. 2013, pp.
  181--188.

\bibitem{Che1}
G.~Zeitler, G.~Kramer, and A.~C. Singer, ``Bayesian parameter estimation using
  single-bit dithered quantization,'' \emph{IEEE Trans. Signal Process.},
  vol.~60, no.~6, pp. 2713--2726, Jun. 2012.

\bibitem{Che2}
J.~Mo, P.~Schniter, and R.~W. Heath~Jr, ``Channel estimation in broadband
  millimeter wave {MIMO} systems with few-bit {ADC}s,'' \emph{IEEE Trans.
  Signal Process., to be published}.

\bibitem{Che3}
C.~K. Wen, C.~J. Wang, S.~Jin, K.~K. Wong, and P.~Ting, ``Bayes-optimal joint
  channel-and-data estimation for massive {MIMO} with low-precision {ADC}s,''
  \emph{IEEE Trans. Signal Process.}, vol.~64, no.~10, pp. 2541--2556, May
  2016.

\bibitem{Che4}
J.~Choi, J.~Mo, and R.~W. Heath, ``Near maximum-likelihood detector and channel
  estimator for uplink multiuser massive {MIMO} systems with one-bit {ADC}s,''
  \emph{IEEE Trans. Commun.}, vol.~64, no.~5, pp. 2005--2018, May 2016.

\bibitem{Li-17TSP}
{Y. Li, C. Tao, G. Seco-Granados, A. Mezghani, A. L. Swindlehurst, and L. Liu},
  ``{Channel estimation and performance analysis of one-bit massive MIMO
  systems},'' \emph{IEEE Trans. Signal Process.}, vol.~65, no.~15, pp.
  4075--4089, Aug. 2017.

\bibitem{Jacobsson-17TWireless}
{S. Jacobsson, G. Durisi, M. Coldrey, U. Gustavsson, and C. Studer},
  ``{Throughput analysis of massive MIMO uplink with low-resolution ADCs},''
  \emph{IEEE Trans. Wireless Commun.}, vol.~16, no.~6, pp. 4038--4051, June
  2017.

\bibitem{Liang-16JSAC}
{N. Liang and W. Zhang}, ``{Mixed-ADC massive MIMO},'' \emph{IEEE J. Sel. Areas
  Commun.}, vol.~34, no.~4, pp. 983--997, Apr. 2016.

\bibitem{LowR1}
\BIBentryALTinterwordspacing
C.~Risi, D.~Persson, and E.~G. Larsson, ``Massive {MIMO} with 1-bit {ADC},''
  2014. [Online]. Available: \url{http://arxiv.org/abs/1404.7736}
\BIBentrySTDinterwordspacing

\bibitem{Liang-16TCOM}
{N. Liang and W. Zhang}, ``{Mixed-ADC massive MIMO uplink in
  frequency-selective channels},'' \emph{IEEE Trans. Commun}, vol.~64, no.~11,
  pp. 4652--4666, Nov. 2016.

\bibitem{det1}
A.~Mezghani, M.~S. Khoufi, and J.~A. Nossek, ``Maximum likelihood detection for
  quantized {MIMO} systems,'' in \emph{Int. ITG Workshop Smart Antennas},
  Vienna, Austria, Feb. 2008, pp. 278--284.

\bibitem{det2}
U.~S. Kamilov, V.~K. Goyal, and S.~Rangan, ``Message-passing de-quantization
  with applications to compressed sensing,'' \emph{IEEE Trans. Signal
  Process.}, vol.~60, no.~12, pp. 6270--6281, Dec. 2012.

\bibitem{det3}
Z.~Wang, H.~Yin, W.~Zhang, and G.~Wei, ``Monobit digital receivers for {QPSK}:
  Design, performance and impact of {IQ} imbalances,'' \emph{IEEE Trans.
  Commun.}, vol.~61, no.~8, pp. 3292--3303, Aug. 2013.

\bibitem{det4}
S.~Wang, Y.~Li, and J.~Wang, ``Multiuser detection in massive spatial
  modulation {MIMO} with low-resolution {ADC}s,'' \emph{IEEE Trans. Wireless
  Commun.}, vol.~14, no.~4, pp. 2156--2168, Apr. 2015.

\bibitem{Dong-17ComLetter}
{Y. Dong and L. Qiu}, ``{Spectral efficiency of massive MIMO systems with
  low-resolution ADCs and MMSE receiver},'' \emph{IEEE Commun. Lett.}, vol.~21,
  no.~8, pp. 1771--1774, Aug. 2017.

\bibitem{Zhang-16ComLetter}
{J. Zhang, L. Dai, S. Sun, and Z. Wang}, ``{On the spectral efficiency of
  massive MIMO systems with low-resolution ADCs},'' \emph{IEEE Commun. Lett.},
  vol.~20, no.~5, pp. 842--845, May 2016.

\bibitem{Kong-17TWireless}
{C. Kong, C. Zhong, S. Jin, S. Yang, H. Lin, and Z. Zhang}, ``{Full-duplex
  massive MIMO relaying systems with low-resolution ADCs},'' \emph{IEEE Trans.
  Wireless Commun.}, vol.~16, no.~8, pp. 5033--5047, Aug. 2017.

\bibitem{Kong-17ArXiv}
\BIBentryALTinterwordspacing
{C. Kong, A. Mezghani, C. Zhong, A. L. Swindlehurst, and Z. Zhang},
  ``{Multipair massive MIMO relaying systems with one-bit ADCs and DACs},''
  Mar. 2017. [Online]. Available: \url{https://arxiv.org/abs/1703.08657}
\BIBentrySTDinterwordspacing

\bibitem{Fan-15ComLetter}
{L. Fan, S. Jin, C.-K. Wen, and H. Zhang}, ``{Uplink achievable rate for
  massive MIMO systems with low-resolution ADC},'' \emph{IEEE Commun. Lett.},
  vol.~19, no.~12, pp. 2186--2189, Dec. 2015.

\bibitem{LowR3}
X.~Zhang, M.~Matthaiou, M.~Coldrey, and E.~Bj{\"o}rnson, ``Impact of residual
  transmit {RF} impairments on training-based {MIMO} systems,'' \emph{IEEE
  Trans. Commun.}, vol.~63, no.~8, pp. 2899--2911, Aug. 2015.

\bibitem{LowR5}
J.~Mo and R.~W. Heath, ``Capacity analysis of one-bit quantized {MIMO} systems
  with transmitter channel state information,'' \emph{IEEE Trans. Signal
  Process.}, vol.~63, no.~20, pp. 5498--5512, Oct. 2015.

\bibitem{SQUID}
S.~Jacobsson, G.~Durisi, M.~Coldrey, T.~Goldstein, and C.~Studer, ``Quantized
  precoding for massive {MU-MIMO},'' \emph{IEEE Trans. Commun.}, vol.~65, no.~11,
  pp. 4670--4684, Nov. 2017.

\bibitem{R1}
------, ``Nonlinear 1-bit precoding for massive {MU-MIMO} with higher-order
  modulation,'' in \emph{Proc. Asilomar Conf. Signals, Syst. and Comput.},
  Pacific Grove, CA, USA, Nov 2016, pp. 763--767.

\bibitem{R2}
\BIBentryALTinterwordspacing
O.~Casta{\~{n}}eda, S.~Jacobsson, G.~Durisi, M.~Coldrey, T.~Goldstein, and
  C.~Studer, ``1-bit massive {MU-MIMO} precoding in {VLSI},'' 2017. [Online].
  Available: \url{http://arxiv.org/abs/1702.03449}
\BIBentrySTDinterwordspacing

\bibitem{R4}
H.~Jedda, J.~A. Nossek, and A.~Mezghani, ``Minimum {BER} precoding in 1-bit
  massive {MIMO} systems,'' in \emph{Proc. IEEE Sensor Array and Multichannel
  Signal Process. Workshop (SAM)}, Rio de Janerio, Brazil, Jul. 2016, pp. 1--5.

\bibitem{R5}
O.~B. Usman, H.~Jedda, A.~Mezghani, and J.~A. Nossek, ``{MMSE} precoder for
  massive {MIMO} using 1-bit quantization,'' in \emph{Proc. IEEE Int. Conf.
  Acoust., Speech, Signal Process. (ICASSP)}, Shanghai, China, Mar. 2016, pp.
  3381--3385.

\bibitem{R6}
\BIBentryALTinterwordspacing
S.~Jacobsson, G.~Durisi, M.~Coldrey, and C.~Studer, ``Massive {MU-MIMO-OFDM}
  downlink with one-bit {DAC}s and linear precoding,'' 2017. [Online].
  Available: \url{http://arxiv.org/abs/1704.04607}
\BIBentrySTDinterwordspacing

\bibitem{A1}
S.~K. Mohammed and E.~G. Larsson, ``Single-user beamforming in largescale
  {MISO} systems with per-antenna constant-envelope constraints: The doughnut
  channel,'' \emph{IEEE Trans. Wireless Commun.}, vol.~11, no.~11, pp.
  3992--4005, Nov. 2012.

\bibitem{am}
------, ``Per-antenna constant envelope precoding for large multi-user {MIMO}
  systems,'' \emph{{IEEE Trans. Commun.}}, vol.~61, no.~3, pp. 1059--1071, Mar.
  2013.

\bibitem{A3}
------, ``Constant-envelope multi-user precoding for frequency-selective
  massive mimo systems,'' \emph{IEEE Wireless Commun. Lett.}, vol.~2, no.~5,
  pp. 547--550, Oct. 2013.

\bibitem{cross_entropy}
{J. C. Chen, C. K. Wen, and K. K. Wong}, ``Improved constant envelope multiuser
  precoding for massive {MIMO} systems,'' \emph{IEEE Commun. Lett.}, vol.~18,
  no.~8, pp. 1311--1314, Aug 2014.

\bibitem{A5}
W.~X. L.~Liang and X.~Dong, ``Low-complexity hybrid precoding in massive
  multiuser mimo systems,'' \emph{IEEE Wireless Commun. Lett.}, vol.~3, no.~6,
  pp. 653--656, Dec. 2014.

\bibitem{A6}
J.~C. Chen, ``Hybrid beamforming with discrete phase shifters for
  millimeter-wave massive mimo systems,'' \emph{IEEE Trans. Veh. Technol.},
  vol.~66, no.~8, pp. 7604--7608, Aug. 2017.

\bibitem{CEP}
{S. K. Mohammed and E. G. Larsson}, ``Per-antenna constant envelope precoding
  for large multi-user {MIMO} systems,'' \emph{IEEE Trans. Commun.}, vol.~61,
  no.~3, pp. 1059--1071, Mar. 2013.

\bibitem{mitchell2002branch}
{P. M. Pardalos and M. G. C. Resende}, \emph{Handbook of applied
  optimization}.\hskip 1em plus 0.5em minus 0.4em\relax Oxford University
  Press, 2000, pp. 65--67.

\bibitem{TB_CEP}
{M. Kazemi, H. Aghaeina and T. M. Duman}, ``Discrete-phase constant envelope
  precoding for massive {MIMO} systems,'' \emph{IEEE Trans. Commun.}, vol.~65,
  no.~5, pp. 2011--2021, May 2017.

\bibitem{WFP}
M.~Joham, W.~Utschick, and J.~A. Nossek, ``Linear transmit processing in {MIMO}
  communications systems,'' \emph{IEEE Trans. Signal Process.}, vol.~53, no.~8,
  Aug. 2005.

\bibitem{CEP2}
A.~Liu and V.~K.~N. Lau, ``Two-stage constant-envelope precoding for low-cost
  massive {MIMO} systems,'' \emph{IEEE Trans. Signal Process.}, vol.~64, no.~2,
  pp. 485--494, Jan. 2016.

\bibitem{NOMAD}
S.~Le~Digabel, ``Algorithm 909': {NOMAD}: Nonlinear optimization with the
  {MADS} algorithm,'' \emph{ACM Trans. Math. Software (TOMS)}, vol.~37, no.~4,
  p.~44, Feb. 2011.

\bibitem{ADMM}
S.~Boyd, N.~Parikh, E.~Chu, B.~Peleato, and J.~Eckstein, ``Distributed
  optimization and statistical learning via the alternating direction method of
  multipliers,'' \emph{Found. and Trends Mach. Learn.}, vol.~3, pp. 1--122,
  2011.

\bibitem{ADMM-infinity-norm}
S.~Shahabuddin, M.~Juntti, and C.~Studer, ``Admm-based infinity norm detection
  for large mu-mimo: Algorithm and vlsi architecture,'' in \emph{Proc. IEEE
  Int. Symp. Circuits and Syst. (ISCAS)}, MD, USA, May 2017, pp. 1--4.

\bibitem{Unbiased}
{S. M. Kay}, \emph{{Fundamentals of Statistical Signal Processing: I.
  Estimation Theory}}.\hskip 1em plus 0.5em minus 0.4em\relax Prentice-Hall
  Inc., Upper Saddle River, NJ, USA, 1993, Ch. 2.

\bibitem{inverse-lemma}
W.~W. Hager, ``Updating the inverse of a matrix,'' \emph{SIAM Rev.}, vol.~31,
  no.~2, pp. 221--239, 1989.

\end{thebibliography}
\end{document}